\documentclass[twocolumn]{aastex63}

\usepackage{xcolor}
\usepackage{amsmath}
\usepackage{affils}

\newcommand{\reff}[0]{${\rm R_{eff}}$}
\newcommand{\sersic}[0]{S\'ersic}
\newcommand{\figdescrip}[0]{\textsf{[fill in figure 
description when figure is finalized]}}
\newcommand{\needscites}[0]{\textsf{[need to fill
in citations here/in introduction]}}
\newcommand{\vmax}[0]{$1/{\rm V_{\max{}}}$}
\newcommand{\thh}[0]{$^{\rm th}$}
\newcommand{\kms}[0]{{\ \rm km\ s}^{-1}}
\newcommand{\song}[1]{\textcolor{blue}{[Song: #1]}}
\newcommand{\rrr}[1]{\textcolor{black}{#1}}

\begin{document}
\shortauthors{Kado-Fong et al.}

\title{Tracing the Intrinsic Shapes of Dwarf Galaxies out to Four Effective Radii: Clues to Low-Mass Stellar Halo Formation}

\newcommand{\princeton}[0]{Department of Astrophysical Sciences, Princeton University,Princeton, NJ 08544, USA}
\DeclareAffil{princeton}{Department of Astrophysical Sciences, Princeton University,Princeton, NJ 08544, USA}
\DeclareAffil{naoj}{National Astronomical Observatory of Japan, 2-21-1 Osawa, Mitaka, Tokyo 181-8588, Japan}
\DeclareAffil{sokendai}{Graduate University for Advanced Studies(SOKENDAI), 2-21-1 Osawa, Mitaka, Tokyo 181-8588, Japan}
\DeclareAffil{hubble}{Hubble Fellow}
\DeclareAffil{carnegie}{The Observatories of the Carnegie Institution for Science, 813 Santa Barbara Street, Pasadena, CA 91101, USA}

\author[ 0000-0002-0332-177X]{Erin Kado-Fong}
\affiliation{Department of Astrophysical Sciences, Princeton University,Princeton, NJ 08544, USA}
\author{Jenny E. Greene}
\affiliation{Department of Astrophysical Sciences, Princeton University,Princeton, NJ 08544, USA}
\author[0000-0003-1385-7591]{Song Huang}
\affiliation{Department of Astrophysical Sciences, Princeton University,Princeton, NJ 08544, USA}
\author[0000-0002-1691-8217]{Rachael Beaton}
\altaffiliation{Hubble Fellow}
\affiliation{Department of Astrophysical Sciences, Princeton University,Princeton, NJ 08544, USA}
\affiliation{The Observatories of the Carnegie Institution for Science, 813 Santa Barbara Street, Pasadena, CA 91101, USA}
\author{Andy D. Goulding}
\affiliation{Department of Astrophysical Sciences, Princeton University,Princeton, NJ 08544, USA}
\author[0000-0002-3852-6329]{Yutaka Komiyama}
\affiliation{National Astronomical Observatory of Japan, 2-21-1 Osawa, Mitaka, Tokyo 181-8588, Japan}
\affiliation{Graduate University for Advanced Studies (SOKENDAI), 2-21-1 Osawa, Mitaka, Tokyo 181-8588, Japan}

\correspondingauthor{Erin Kado-Fong} 
\email{kadofong@princeton.edu}

  
\date{\today}

\begin{abstract}
Though smooth, extended spheroidal stellar outskirts have
long been observed around nearby dwarf galaxies, it is unclear whether
dwarfs generically host an extended stellar halo. We use imaging from
the Hyper Suprime-Cam Subaru Strategic Program (HSC-SSP) to 
measure the shapes of dwarf galaxies out to four effective radii for 
a sample of dwarfs at $0.005<z<0.2$ and $10^{7.0}<M_\star/M_\odot<10^{9.6}$.
We find that dwarfs are slightly triaxial, 
with a $\langle B/A \rangle \gtrsim0.75$ 
(where the ellipsoid is characterized by three principle semi-axes 
constrained by $C\leq B\leq A$). At $M_\star>10^{8.5}M_\odot$, the 
galaxies grow from thick disk-like near their centers towards 
the spheroidal extreme at four effective radii. We also see that although blue dwarfs
are, on average, characterized by thinner discs than red dwarfs, both blue
and red dwarfs grow more spheroidal as a function of radius. This relation
also holds true for a comparison between field and satellite dwarfs.
This uniform trend towards relatively spheroidal shapes as a function
of radius is consistent with an in-situ formation mechanism for stellar outskirts
around low-mass galaxies, in agreement with proposed models where
star formation feedback produces round stellar outskirts around dwarfs.
\end{abstract}

\section{Introduction}
The existence of a smooth stellar component in the outskirts 
of local dwarfs is a common, but puzzling, phenomenon \citep{lin1983,minniti1996,grebel1999,minniti1999,roychowdhury2013}. 
Round stellar halos are a near-ubiquitous
component of more massive galaxies -- thought to be assembled largely
through the accretion of satellite galaxies, the stars that populate these
outskirts provide key insights into the galaxy's assembly history 
\citep[see, e.g.][]{bullock2005, abadi2006}. 
Due to a decreasing stellar mass to halo mass ratio, satellite accretion
by dwarf centrals deposits fewer stars per unit halo mass than 
analogous events around more massive systems \citep{purcell2007,brook2014}.
It is thus considered unlikely that minor mergers are able to fuel
the formation of a stellar halo in dwarf galaxies.

{}

Instead, it has been suggested that the
stellar outskirts of dwarfs are an in-situ structure.
In the field, dwarf galaxies sit in shallow potential wells; their structure is therefore
more sensitive to the details of star formation feedback than more massive
galaxies. 
Supernovae-driven winds \citep{hu2019}, cosmic ray feedback \citep{dashyan2020},
stellar winds, radiation pressure, and photoionization \citep{elbadry2016} are all expected
to more efficiently displace gas in dwarfs than in more 
massive hosts (both in moving gas to large radii and in
removing it from the system entirely).
In particular, hydrodynamical simulations have predicted that star formation
feedback can induce significant size fluctuations in the stellar content of dwarf galaxies,
 driving the formation of a round
stellar halo by inducing radial migration via potential fluctuations, as well as 
forming stars in outflowing and inflowing gas \citep{stinson2009,maxwell2012, elbadry2016}.

Not all theories of dwarf stellar halo formation are purely in-situ, 
however; \citet{bekki2008} suggested that round stellar outskirts
around dwarfs may be 
formed as a product of dwarf-dwarf major mergers (a merger wherein the mass of the
secondary is at most a factor of $\sim3$ 
less than that of the primary). Such major mergers are
expected to occur for about 75\% of galaxies in this stellar mass
range, but are expected to proceed far more often in the early universe
-- only 30\% of these galaxies are expected to have undergone a 
major merger in the last 10 Gyr
\citep{deason2014,besla2018}. Stellar outskirts formed in this
manner would tend to be comprised of ancient stellar populations;
using the surface brightness profile given by \citet{bekki2008} 
directly after the outskirts are formed and taking into
account surface brightness dimming due to passive evolution 
\citep{conroy2009}, in the major merger scenario 
we would not expect to detect an extended round stellar component 
around the majority of dwarfs.


Moreover, the dwarfs that
have been found to host extended, smooth 
intermediate-old age stellar populations are nearby systems in the 
Local Volume \citep[and mostly in the Local Group, see][]{zaritsky2000,aparicio2000,aparicio2000a,hidalgo2003, demers2006,bernard2007,stinson2009,strader2012,nidever2019a,nidever2019,pucha2019}. 
It remains unclear
whether such a structure is a generic feature
of dwarfs (pointing to an in-situ origin), or a
result of the influence of the more massive galaxies
in the Local Group. 


Understanding the intrinsic shape of dwarf galaxies is thus of
interest in understanding the stellar assembly of these low-mass systems, and for
constraining recipes for star formation feedback.
However, it has
historically been challenging to construct a sample of dwarfs with sufficient
numbers whose imaging is deep enough to measure stable ellipticity profiles.
Previous works have been confined to the Local Volume \citep{roychowdhury2013},
 or to the most massive dwarfs 
 \citep[$M_\star>10^9M_\odot$,][]{padilla2008, vanderwel2014, zhang2019}. 
Moreover, there has not been an effort previously to measure the intrinsic
shapes of dwarf outskirts, due largely to the aforementioned
technical hurdles.

In this work, we combine the large sample of spectroscopically 
confirmed dwarfs observed by the Sloan Digital Sky Survey spectroscopic
surveys 
\citep[both legacy and BOSS surveys,][]{strauss2002, dawson2013, reid2016} 
and the Galaxy and Mass Assembly (GAMA) spectroscopic 
survey \citep{baldry2012} with the wide and deep imaging of the
Hyper Suprime-Cam Subaru Strategic
Program \citep[HSC-SSP;][]{HSC1stDR,HSC1styrOverview,Miyazaki18HSC,komiyama2018,Kawanomoto18HSC,Furusawa18HSC,Bosch18HSC,Haung18HSC,Coupon18HSC} to quantify the 3D shape distribution
of dwarf galaxies as a function of radius.

The wide area covered by the HSC-SSP in conjunction with the
surface brightness sensitivity and high resolution of its imaging allows
us to map stable ellipticity profiles of the dwarfs out to four times
the half-light radius ($R=4$\reff{}, where \reff{} is defined by a 
single \sersic{} fit, as described in \autoref{s:1dprofiles})
at $0.005<z<0.2$ and $7.0\leq \log_{10}(M_\star/M_\odot)\leq9.6$. 
This allows us to construct a sufficiently large sample of dwarfs to infer
the distribution of their intrinsic shapes from observations of their
projected 2D shapes at fixed radius. In \autoref{s:observations}, we detail the
sample selection and volume corrections implemented for the sample.
We detail the methodology and validation of the 1D surface brightness
profiles and the 3D shape inference separately, in \autoref{s:1dprofiles}
and \autoref{s:3dshapeinference}, respectively. We then examine
the change in dwarf 3D shape as a function of radius and dwarf
properties in \autoref{s:results}, and consider the implications of the
observed shape evolution to proposed dwarf stellar halo formation
mechanisms in \autoref{s:discussion}.

Throughout this paper we adopt a standard flat $\Lambda$CDM model 
in which H$_0=70$ km s$^{-1}$ Mpc$^{-1}$ and $\Omega_m=0.3$.

\begin{figure}[htb]
\includegraphics[width=\linewidth]{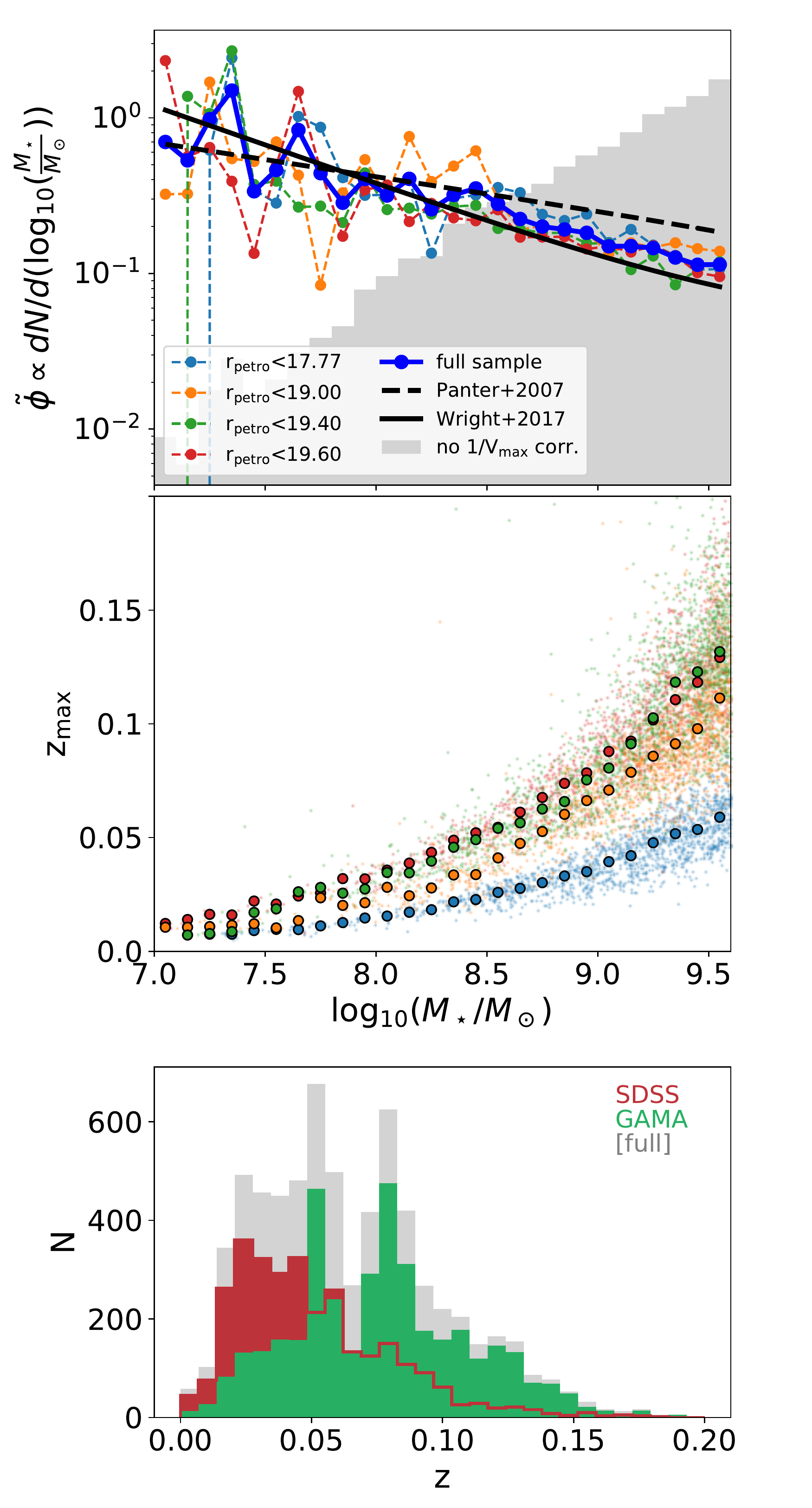}
\caption{ 
    \textit{Top:} we show the normalized stellar mass function (SMF),  
    $\tilde \phi (\log_{10}(M_\star/M_\odot))$, for each magnitude-limited
    subset of our sample with dash-connected scatter. The normalized SMF for the
    full sample is shown by the solid blue curve. For reference, the 
    Schechter fit of \citet{panter2007} and the double Schechter fit of 
    \citet{wright2017} are normalized to our stellar mass range and
    shown in the dashed and solid lines, respectively. The raw distribution in
    stellar mass (i.e. without \vmax{} weights) is shown by the grey filled
    histogram. We find that our normalized SMF matches that of \citet{wright2017}
    well. \textit{Middle:} the maximum observable redshift, $z_{\rm max}$, for each
    galaxy in the sample as a function of stellar mass. 
    As at top, points are colored by the magnitude limit of the 
    source spectroscopic program.
    The mean $z_{\rm max}$ as a function of stellar mass is also shown for each
    program by the large scatter points.
    \textit{Bottom:} the redshift distribution, colored by source survey. 
    }
\label{f:vmaxcorrection}
\end{figure}

\section{Observations and Data Processing}\label{s:observations}
\subsection{HSC-SSP Imaging}
As noted above, the HSC-SSP imaging boasts wide, deep, and
high resolution imaging, making it well-suited for an exploration of
the low surface brightness outskirts of low-mass galaxies. 
Upon completion, HSC-SSP will provide imaging with a median seeing of $\sim0.6''$ in the 
$i_{\rm HSC}$ over $\sim 1400$ square degrees to a point source
depth of $i_{\rm HSC}\sim 26$ in its shallowest ``Wide'' 
layer \citep{aihara2019}. The data have been shown to reach
surface brightness limits of $i_{\rm HSC}\sim28.5$ mag arcsec$^{-2}$ 
for measurements around a known target \citep{huang2018}. 
\rrr{We test the surface brightness limit of the HSC-SSP data in the 
vicinity of our sample in \autoref{s:backgroundsubtraction}, and find
that $\mu_i=28.5$ mag arcsec$^{-2}$ is a conservative choice of limiting 
surface brightness. Indeed, with an empirical correction to the background, it has been shown that 
HSC-SSP reaches depths of $\mu_r\sim 29.5$ mag arcsec$^{-2}$. However, because
we allow several parameters to drift during our surface brightness profile measurements,
we adopt the fiducial surface brightness limit of $\mu_i=28.5$ mag arcsec$^{-2}$.}

For this work, we use the \rrr{internal} HSC\rrr{-SSP} S18A data release, which 
covers the same area as the second public data release of 
\citet{aihara2019} \rrr{and is processed with a very similar 
data reduction pipeline. Though are there some minor differences between the
data reduction pipeline used for S18A and PDR2, these changes do not affect
the parts of the pipeline discussed in this work.} We require only coverage in
the $i_{\rm HSC}$ band, resulting in an area of $\sim 796$ 
deg$^2$ with a point source depth of $i=26.2^{+0.2}_{-0.4}$.
\rrr{The $i_{\rm HSC}$ band is best choice to study the overall
shape of the stellar distribution for two main 
reasons. First, the $i_{\rm HSC}$ bands has best seeing out of the 
five HSC bands. Second, it is less sensitive to dust extinction and star
forming regions relative to the bluer $g_{\rm HSC}$ and 
$r_{\rm HSC}$ bands, and deeper than the redder $z_{\rm HSC}$ and
$y_{\rm HSC}$ are significantly shallower.}

\subsection{Initial Sample Selection}
All dwarfs in the present sample have been spectroscopically 
observed by either the SDSS or GAMA spectroscopic surveys.
We limit our sample to dwarfs at $0.005<z<0.2$; the redshift distribution of the sample peaks at $z\lesssim0.05$, as shown in the bottom panel  
of \autoref{f:vmaxcorrection}. Due to the intrinsic faintness of low-mass galaxies,
the majority of our sample is at $z\lesssim 0.1$.

This selection yields a sample of 11338 dwarfs. In 3128 cases, there is a
bright star (or related imaging artifact) within 5\reff{} of the 
target galaxy; because our goal is to measure 
ellipticity profiles out to the outskirts of the dwarfs, we remove these
galaxies from the final sample. An additional 548 galaxies are too close to
neighbors to measure a reliable surface brightness profile, or are coincident
with an imaging artifact. 

We adopt stellar masses measured by the SDSS and GAMA teams. The stellar 
masses measured by the GAMA team use a Chabrier initial
mass function \citep{chabrier2003} by \citet{taylor2011}. The stellar masses
measured by the SDSS team are derived using the \citet{conroy2009} 
Flexible Stellar Population Synthesis (FSPS) models with a Kroupa
initial mass function \citep{kroupa2001}. In \citet{kadofong2020}, we found
that, for galaxies with both SDSS and GAMA spectroscopy,
the SDSS stellar masses are higher than the GAMA stellar masses by a
median of 0.08 dex and a median absolute deviation of 0.35 dex. We
therefore reduce the masses derived from SDSS observations by 0.08 dex; 
it is however important to note that due to the width of our mass bins,
including or excluding this shift does not impact this work.

\subsection{Volume Corrections}
Our sample is drawn from the SDSS and GAMA spectroscopic surveys, both
of which comprise several subsets with different magnitude limits.
The sample is composed of observations from the SDSS Legacy 
Survey \citep[$r_{\rm petro} < 17.77$, ][]{strauss2002}, 
the low redshift component of SDSS BOSS \citep[$r_{\rm petro} < 19.6$, ][]{dawson2013},
and the GAMA second public release \citep[$r_{\rm petro} < 19.4$ or $r_{\rm petro} < 19.0$,
depending on the region; see][]{liske2015}. 
To convert this sample from magnitude-limited to volume-limited, 
we adopt the classical \vmax{} correction to simulate a volume-limited 
sample.

\begin{figure}[htb]
\centering     
\includegraphics[width=\linewidth]{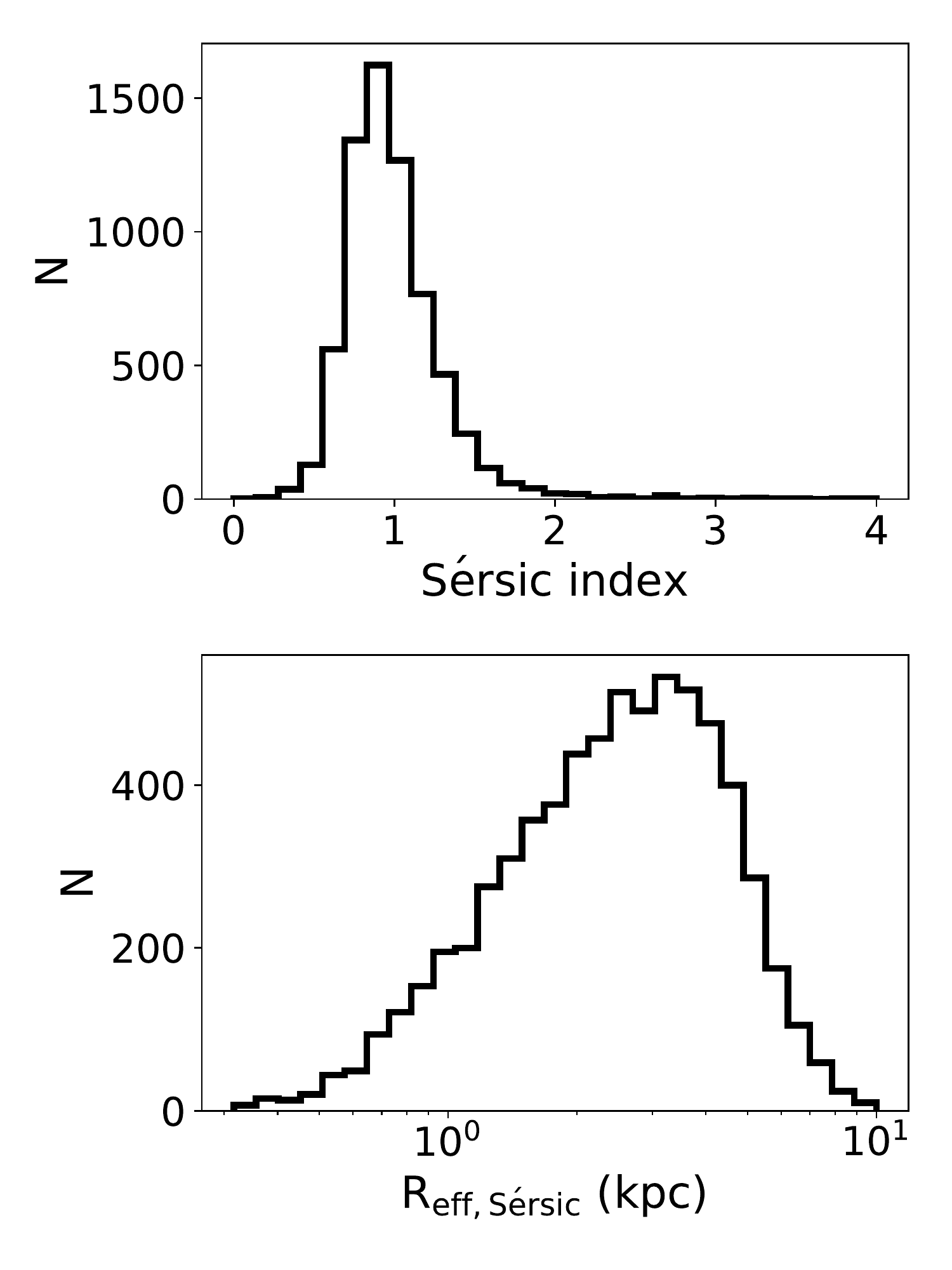}
\caption{ 
 \rrr{\textit{Top}: The distribution of
    \sersic{} indices over the dwarf sample. The majority of the dwarfs are well-described
    by an exponential ($n=1$) profile. \textit{Bottom}: The distribution of effective
    radii in the sample, as measured from the \sersic{} fits.}
    }
\label{f:sersicfits}
\end{figure}  

\begin{figure*}[htb]
\centering     
\includegraphics[width=\linewidth]{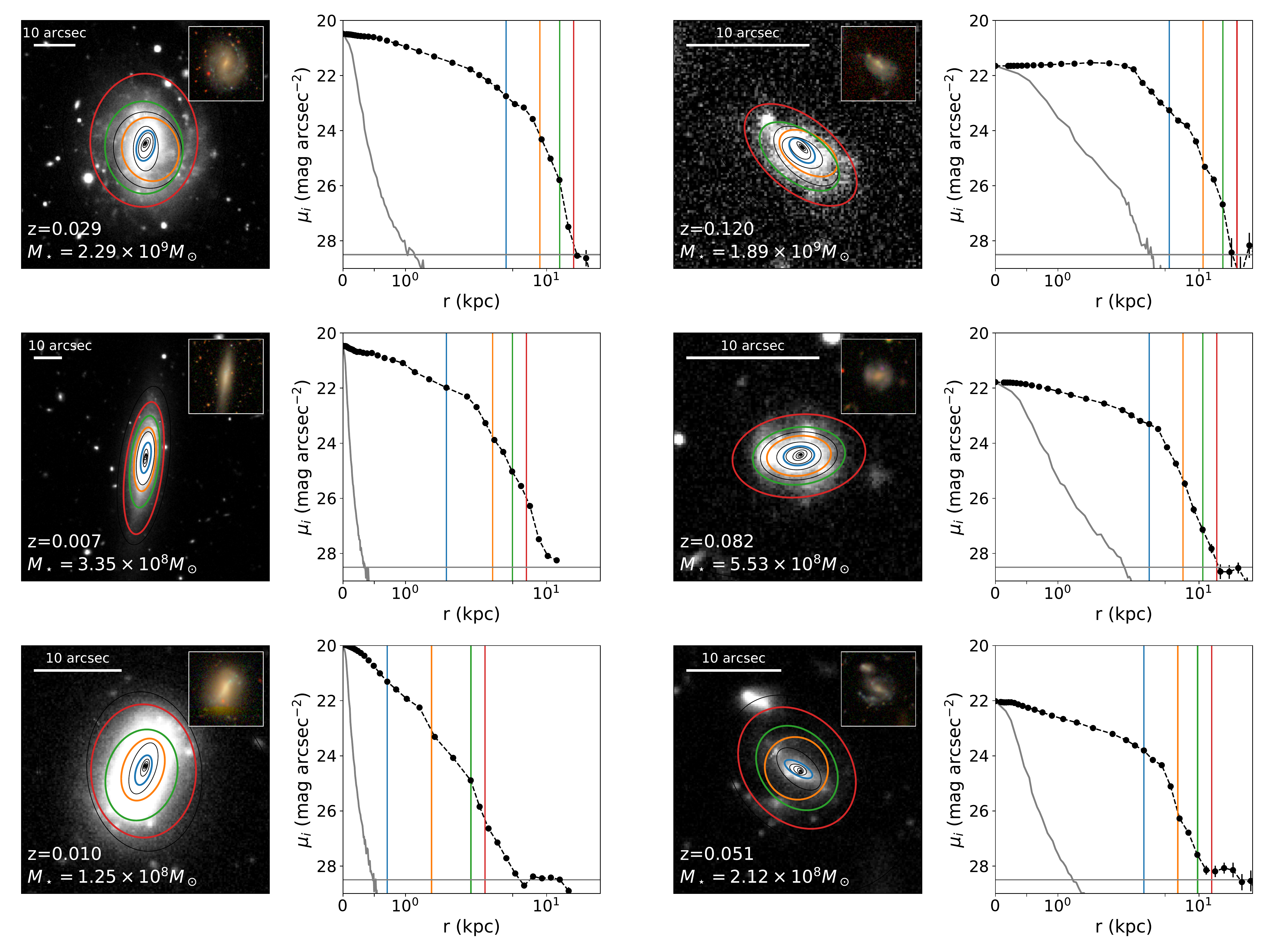} 
\caption{
   Example 1D surface brightness profiles for galaxies that span the 
   range of stellar mass and redshift
   in our sample. The lefthand (righthand) columns show galaxies
   at the low (high) redshift end of the sample, while the rows are ordered
   in decreasing stellar mass. Each pair of panels shows the 
   $i_{\rm HSC}$-band image with the measurements at 1\reff{} (blue), 
   2\reff{} (orange), 3\reff{} (green), and 4\reff{} (red) overplotted as
    ellipses. The $gri_{\rm HSC}$-composite RGB image 
    is also shown in the inset panel to more
   clearly show the morphology of the galaxy. The right panel shows the
   1D surface brightness profile of the galaxy; vertical lines show the physical
   extent at 1-4\reff{} (same colors as left). We also show the surface brightness
   of the PSF by the grey curve, and the nominal surface brightness limit of
   $\mu_i=28.5$ mag arcsec$^{-2}$ by the grey horizontal line. 
    }
\label{f:sbprofiles}
\end{figure*}

As all of the galaxies in our sample are low mass, none have maximum observable redshifts
for which the observed-frame $r$-band lies outside of the wavelength range of the
(SDSS or GAMA) optical spectrograph. We are thus able to compute the maximum redshift at which
the observed r$_{\rm SDSS}$ Petrosian magnitude lies within the spectroscopic selection,
$z_{\max}$, directly
from the spectra using the public filter response curves measured for SDSS in 
2001\footnote{http://www.sdss3.org/instruments/camera.php}. We use the catalog SDSS and GAMA
r$_{\rm SDSS}$ Petrosian magnitudes to compute $z_{\max{}}$.
Though the spectroscopic
observations span a considerable range in time, it has been shown that the SDSS $r$-band 
filter transmission curve has evolved by less than 0.01 mag \citep{doi2010}. We therefore
use the fiducial SDSS transmission curve for all galaxies. 

We remove galaxies for which $z \geq z_{\max{}} + 0.005$ or 
$z_{\max{}} < 0.005$ (recall that our minimum redshift cut is $z=0.005$), as these
conditions suggest that there is a problem with the catalog photometry or spectroscopy. From
inspection, these are largely comprised of cases where a large galaxy has been erroneously divided into several ``low-mass galaxies'' during
image segmentation \citep[i.e.\ shredding, see][]{blanton2011}.

To validate our directly computed $z_{\max{}}$ values, we compare the distribution of
stellar masses in our sample, as weighted by \vmax{}, to published stellar mass functions
in the literature. We report the stellar mass function normalized over our sample mass range;
that is,
$\tilde\phi(\log_{10}(M_\star/M_\odot))\equiv C_0 dN/d(\log_{10}(M_\star/M_\odot))$
 where $C_0$ is a constant defined such that
 $\int_{7.0}^{9.6}\tilde\phi(x)dx = 1$.

The top panel of \autoref{f:vmaxcorrection} shows this normalized stellar mass function
of each magnitude-limited subset in our sample as dashed-line curves. The distribution
over the full sample is shown by the thick blue lines, while the original unweighted
stellar mass distribution of the sample is shown by the filled grey histogram. The double Schechter fit of \citet[][from GAMA]{wright2017} is shown by
the solid black line, and the Schechter fit of \citet[][from SDSS]{panter2007}
by the dashed black line.
In both cases, the parametric fits are normalized over the stellar mass range of
our sample. 

Our \vmax-corrected stellar mass distribution is in good agreement with the results of
\citet{wright2017}, and is somewhat steeper than the mass function of \citet{panter2007}.
This is expected, as \citet{wright2017} includes significantly more galaxies at the
stellar mass range of the present sample; the agreement between our normalized stellar
mass function and that of \citet{wright2017} indicates that the \vmax{} weights we
implement are well-behaved.

\section{1D Surface Brightness Profile Measurement}\label{s:1dprofiles}
With a volume-corrected sample in hand, we now turn to the main objective of this
work. The measurement of a 3D shape distribution requires both careful measurements
of the projected 1D surface brightness profiles and a framework with which the
3D shape distribution may be inferred from these projected profiles. 


We first address our adopted 1D profile measurement scheme. 
To establish a reasonable initial guess for the
centroid position, mean ellipticity, and position angle of
the source, we first fit each galaxy
with a single \sersic{} profile. We also use the 
\sersic{} profile fit to measure an effective radius (\reff{}) for
each source. We then extract a 1D ellipticity profile from
each galaxy by allowing the isophotal shape to vary 
with radius.

\begin{figure}[htb]
\centering     
\includegraphics[width=\linewidth]{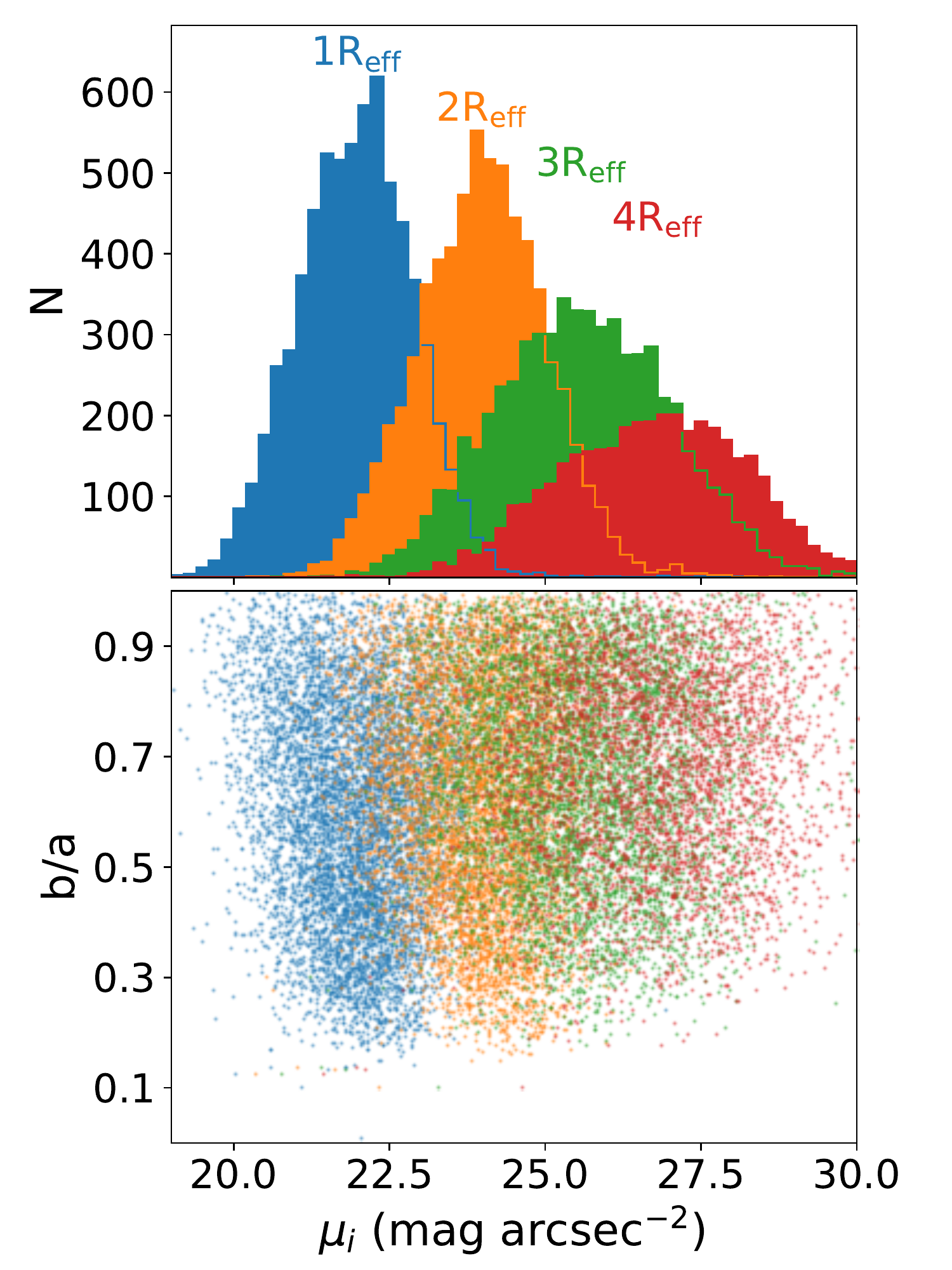}
\caption{ 
    \textit{Top:} the distribution of surface brightness measured at
    $1-4$\reff{}, as labeled. \textit{\rrr{Bottom}:} surface brightness versus 
    projected axis ratio, again colored by measurement radius. Though some
    measurements are below our nominal surface brightness limit of $\mu_i=28.5$
    mag arcsec$^{-2}$, these comprise only $\sim 7\%$ of ellipticity measurements,
    and their inclusion does not have a statistically significant effect on
    the overall ellipticity distribution.}
\label{f:sbpdist}
\end{figure}

\subsection{Single \sersic{} fits}
Though a single \sersic{} model is not flexible enough to fully describe the structure of dwarf galaxies, it provides a stable and robust model to extract basic flux-weighted structural parameters.

The position angle, ellipticity, centroid from this single \sersic{} fit
are used to initialize the non-parametric
surface brightness profile measurement at \reff{}. Because dwarf galaxies 
are often characterized by irregular, off-center star forming regions \citep{binney2008},
an inflexible and monotonically decreasing model is necessary to establish a 
reliable galaxy centroid. 
We additionally adopt the
\sersic-derived effective radius as \reff{} throughout the paper. \rrr{In
\autoref{f:sersicfits}, we show the distribution over \sersic{} index (top) and
 effective radius (bottom) for our
sample. The dwarfs tend to be well-described by an exponential ($n=1$) profile, with a median 
[25\thh{}, 75\thh{} percentile] \sersic{} index of 0.94 [0.79,1.1]. Their effective radii are 
typically a few kpc, with a median [25\thh{}, 75\thh{} percentile] value of 2.5 kpc [1.6, 3.7 kpc],
though we note that there is a strong relationship between stellar mass and
effective radius.}

\begin{figure*}[htb]
\centering     
\includegraphics[width=\linewidth]{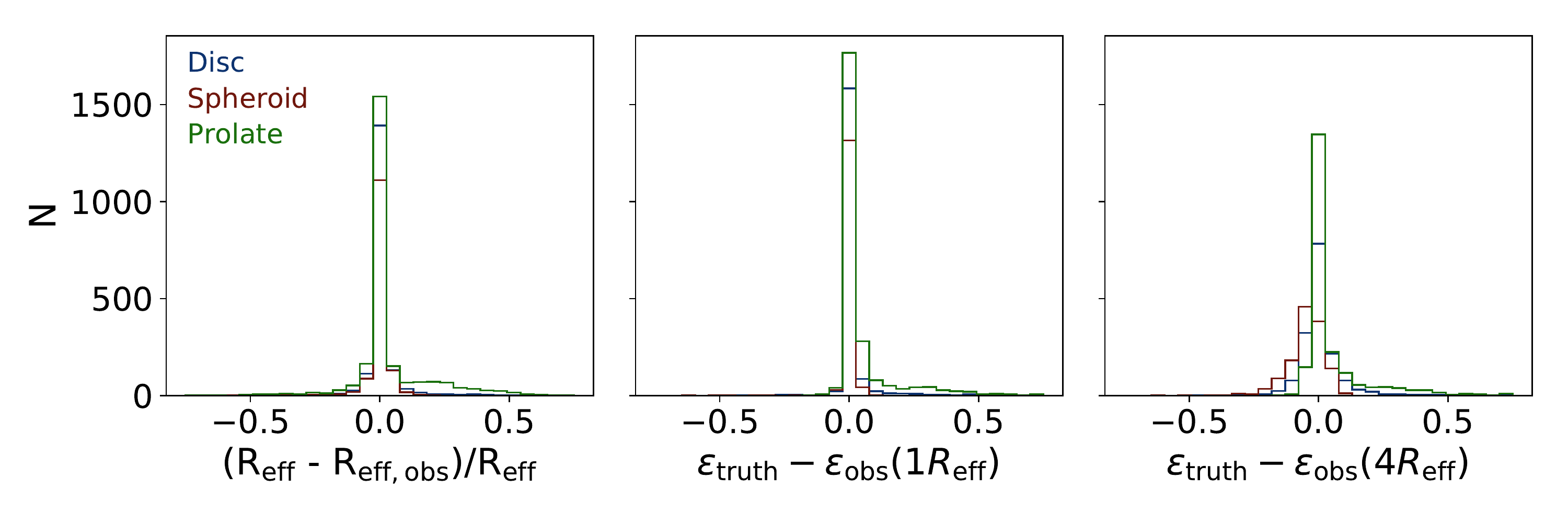} 
\caption{
    The distribution of errors in recovered effective radius (left),
    ellipticity at 1\reff{} (middle), and ellipticity at 3\reff{} (right) 
    for an injected population of disk ($[A,B,C]=[1.,0.9,0.1]$, blue), 
    spheroid ($[A,B,C]=[1.,0.9,0.9]$, red), 
    and prolate ($[A,B,C]=[1.,0.1,0.1]$, green)
    galaxies. Each mock galaxy is assigned a viewing angle drawn isotropically
    over the sphere, injected into the HSC data with an $n=1$ \sersic{} profile,
    and recovered with our pipeline.
    In each panel, the text shows the 5\thh{}, 25\thh{}, 50\thh{},
    75\thh{}, and 95\thh{} percentile of the distribution.
    }
\label{f:extremerecovery}
\end{figure*}

\begin{figure*}[htb]
\centering     
\includegraphics[width=\linewidth]{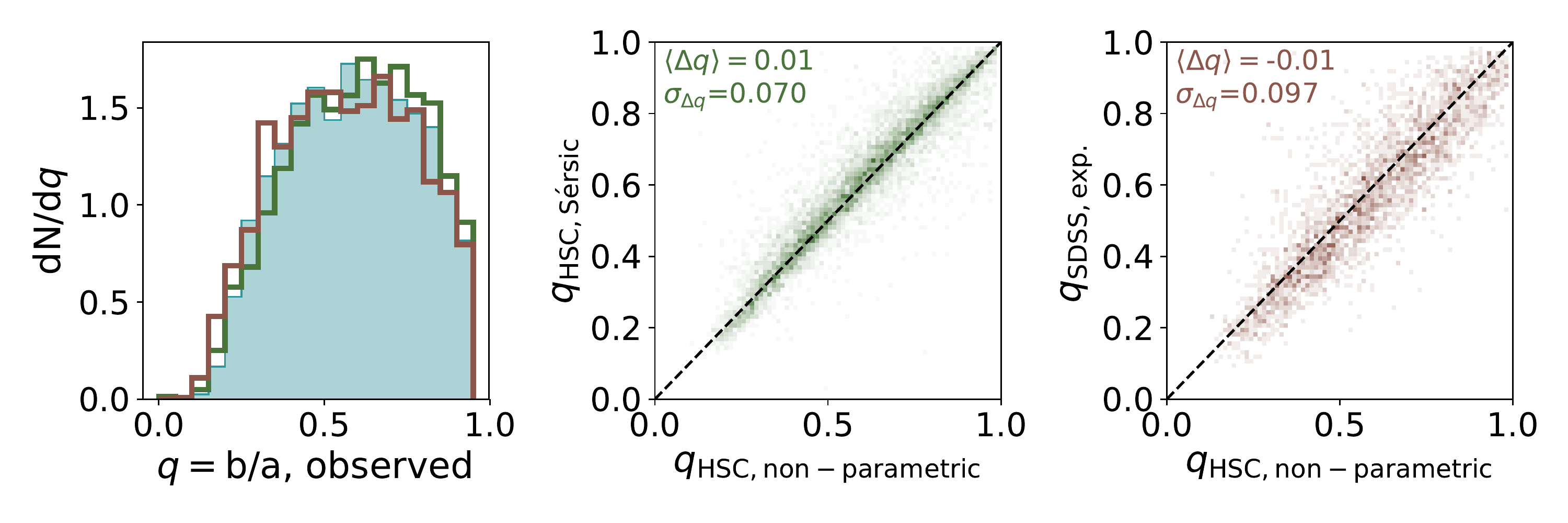} 
\caption{
    \textit{Left:} the observed distribution of projected axis ratio, $q=b/a$,
    for the HSC non-parametric measurements (filled teal histogram), 
    HSC \sersic{} measurements (green histogram), and SDSS catalog measurements
    (derived from exponential fits, brown histogram). All three measurements
    are in good agreement; we find that the mean difference in measured $q$ 
    ($\langle \Delta q \rangle$)
    and the standard deviation of this difference ($\sigma_{\Delta q}$) are
    $(\langle \Delta q \rangle,\sigma_{\Delta q})=(0.017,0.71)$ and
    $(\langle \Delta q \rangle,\sigma_{\Delta q})=(-0.01,0.10)$ for the
    HSC \sersic{} measurements and SDSS catalog exponential measurements, 
    respectively. In the middle panel we plot the non-parametric
    measurements at 1\reff{} against the HSC \sersic{} measurements, 
    while in the right panel we show the same non-parametric measurements
    at 1\reff{} against the SDSS exponential measurements, for galaxies
    in the SDSS catalog.
    }
\label{f:basanity}
\end{figure*}

\subsection{Non-parametric Surface Brightness Profiles}
Though a single \sersic{} fit provides a reasonable initial guess for their 
surface brightness profiles, dwarf galaxies are rich in substructure,
and not well-described by a single \sersic{} profile. Single \sersic{} profiles are
also unable to trace changes in ellipticity as a function of radius, by definition.

In order to better describe the complex structure of low mass galaxies, and
to test whether ellipticity changes as a function of radius, we adopt a more flexible
non-parametric method to measure the 1D surface brightness profiles of the galaxies in
our sample. We use the method introduced by \citet{huang2018}, which is 
based on the \textsf{IRAF} \textsf{Ellipse} algorithm (\citealt{jedrzejewski1987J}), and allows 
ellipticity $\epsilon$, central position $(x_c,y_c)$\footnote{We allow the centroid to drift in order to accommodate the presence of off-center starforming regions that may dominate the light near the center of the galaxy. At the radii we consider for this work, the ellipticity distribution does not change significantly when the centroid is held constant or left free.}, and position angle ${\rm PA}$ to 
vary as a function of semi-major axis $a$.
The exceptional depth of the HSC-SSP imaging allows us to fit profiles with
these parameters free without the fit becoming unstable. 
To further safeguard against an unstable fit, if the centroid shifts by more than
0.5\reff{} at $r={\rm R_{shift}}$, we disregard the surface brightness profile
at $r>{\rm R_{shift}}$.

To generate reliable surface brightness profiles, 
we must first mask out galaxies that are near
the target. To do so, we use the method introduced in 
\citet[][Appendix B]{kadofong2020} 
to detect and mask background sources by detecting sources 
at spatial frequencies that are
high relative to the smooth light of the target outskirts. 
This approach allows us to remove
background galaxies that are at small projected distances from the target galaxy,
where they are most likely to contaminate measurements of the galaxy outskirts. We also apply a $3\sigma$ clipping to the
pixel values along each isophote to reduce the impact
from other objects. \citet{huang2018} and Ardila et al. (in prep.)
have shown that this method works well even for massive
galaxies with extended stellar halos.  We also adopt a 
moderately large
multiplicative step size of $(a_{n+1} - a_n)/a_n = 0.2$ to
help stabilize the ellipticity measurement in the outskirts
of the galaxy.
{}
In \autoref{f:sbprofiles},
we show example 1D profiles that span the stellar mass and redshift range of our
sample. The left panel of each pair shows the isophote at $1-4$\reff{} plotted
over the $i_{\rm HSC}$-band image. The $gri_{\rm HSC}$-composite RGB image is also
shown by the inset panel. The right panel shows the 1D surface brightness profile
for each example. The examples decrease in stellar mass from top to bottom, and
increase in redshift from left to right. In addition, we show the overall distribution
of surface brightness at 1-4\reff{} for our 1D profile fits in 
the top panel of \autoref{f:sbpdist}, and the surface brightness versus ellipticity
in the bottom panel. \rrr{Though we use effective radii (and
multiples thereof) in this work, we have also verified that using 
fixed physical radii does not change our results.}

At 4\reff{}, the farthest extent to which we measure 
ellipticity profiles, the median surface brightness is $\langle\mu_i\rangle=26.8$
mag arcsec$^{-2}$, significantly brighter than our surface brightness limit.
Seven percent of galaxies have a surface brightness of $>28.5$ mag arcsec$^{-2}$ 
at 4\reff{}; we do not remove them from the sample, as their inclusion or exclusion
from this analysis does not have a statistically significant impact on the
overall ellipticity distribution of the overall sample or subsamples considered in 
this work. \rrr{To ensure that we are able to reach this nominal surface brightness
limit, we examine the residual sky background near our dwarf sample in 
\autoref{s:backgroundsubtraction}, and find that the sky is slightly uniformly undersubtracted
near our dwarfs, corresponding to a surface brightness difference of $\Delta \mu_i \lesssim 0.02$
mag arcsec$^{-2}$ at $\mu_i = 28.5$ mag arcsec$^{-2}$. We thus confirm that 
the profiles are well-recovered down to our nominal surface brightness limit, and that
the residual sky does not affect the shape measurements made in the outskirts of the galaxies.}

\subsubsection{The Impact of the Point Spread Function}
In many situations, it is important to correct for the effects of the 
point spread function (PSF) in order to probe the outskirts of galaxies 
\citep[see, e.g.][]{trujillo2016}.
In this work, we expect
that the effect of the PSF does not significantly affect our results for 
the following reasons.
 
First, due to the high resolution of the
HSC imaging, the region of interest for our sample ($r>$\reff{}) is not strongly
affected by the smearing effect of the finite seeing (the median seeing is 
$i_{\rm HSC}\sim0.6$, \citealt{aihara2019}), even at the high redshift end of
our sample. 
To visually demonstrate the size of the PSF with respect to the 
scale of the profile measurements, in \autoref{f:sbprofiles} we plot the
profile of the PSF in grey. 
Second, it is important to note that because 
the cores of dwarfs are intrinsically fainter than those of their more massive 
analogs, the effect of scattered light is smaller for a fixed surface brightness
limit. The theoretical HSC PSF model
of \citet{Coupon18HSC}, which was derived via optical tracing of the 
instrumental and atmospheric PSF, also shows that the large angular-scale
wing of the HSC PSF should not affect measurements made in the outskirts
for a sample such as the one considered in this work. 

To affirm the above statements quantitatively, we apply a series of 
tests to the sample. 
First, we find that the ellipticity distribution (at all radii) does not 
change significantly as a function of redshift; 
one expects that the effect of the PSF would become more pronounced
for higher redshift targets. Second, we re-fit our galaxies with the single
\sersic{} model described in \autoref{s:sersicrecovery_2d} convolved with 
the PSF produced from the HSC-SSP data
reduction pipeline. This is a standard technique used to account for the
effect of the PSF when using a parametric model \citep[for a review, see][]{Knapen2017}. We find that
the ellipticity as measured with this convolved \sersic{} model
and the non-parametric ellipticity measurements at 1\reff{}
are offset by $\mu_{\Delta \epsilon}=-0.030\pm0.001$,
where
$\Delta\epsilon=\epsilon_{\rm 1R_{\rm eff}} - \epsilon_{\rm conv}$. This
shift is statistically significant, but not large enough to 
impact our results.

This is not to say that the effect of the PSF is
generically unimportant for low surface brightness science in HSC images -- 
at $r\gtrsim 30$ kpc and $z\sim0.1$, 
\cite{wang2019} found that the extended HSC PSF can account for up
to 40\% of the flux in the stacked profile of galaxies
at $10^{9.2} < M_\star/M_\odot < 10^{9.9}$ in HSC-SSP. At $r\lesssim20$kpc
with the bulk of the sample sitting at $z<0.1$, 
however, the PSF has a much less significant effect. 
\rrr{In particular, Figure 9 in \cite{wang2019} 
shows that the impact of the PSF, which depends both on the central surface brightness and concentration of the galaxy, should be small for our sample, which is populated by dwarf galaxies with low central surface brightnesses (relative to massive galaxies) and 
exponential profiles. For a dwarf with an exponential
profile and a central surface brightness of
$\mu_i\sim20$ mag arcsec$^{-2}$, the extended wings of 
the HSC PSF appear at around a scale of 2 arcsec and
a surface brightness of 27 mag arcsec$^{-2}$. At scales comparable to 4\reff{}, this effect will be even smaller. We thus do not expect that the PSF will affect 
the results presented in this work.}

\subsection{Profile Measurement Validation}
In order to test the validity of our inferred size and ellipticity profiles,
we perform several mock galaxy injection tests. In particular, we first
confirm that we can recover the ellipticity of \sersic{} profiles injected 
at low surface brightness without significant bias. Then, we compare our 
non-parameteric measurements of ellipticity at 1\reff{} to both the
parametric measurements of our \sersic{} fits to the HSC data and published
ellipticity measurements from SDSS imaging to verify that there is no
systematic shift between the parametric and non-parametric measurements.

\subsubsection{Recovery of Injected \sersic{}s}\label{s:sersicrecovery_2d}
First, we inject mock galaxies composed of a single \sersic{} profile, the 
parameters of which (effective surface brightness, effective radius, 
and intrinsic axis ratios) are drawn 
from a known distribution, into the HSC co-adds.
The surface brightness of the mock galaxies is
set to cover the same range as the real galaxies,
and the \sersic{} index is fixed at $n=1$ 
(exponential) for all galaxies.
The position of the mock galaxies are selected to be
empty locations where there are no detections in any of 
the five HSC bands, 
but no other constraints are made on the galaxy placement. 
These mock galaxies should therefore be affected by
imaging artifacts, background galaxies, and residual 
astrophysical fore/background (e.g. galactic cirrus) in the same way as are the
real galaxies in our sample.

\autoref{f:extremerecovery} shows the injected and 
recovered distributions of the
galaxy structural parameters. We find that the properties of the injected galaxies 
in this simple test are well-recovered: at 4\reff{}, for 75\% of
cases the ellipticity of the
mock galaxy is recovered to better than $|\epsilon_{\rm truth} - \epsilon_{\rm obs}| \leq 0.065$. 
Crucially, decreasing surface brightness 
does not significantly bias the our ellipticity 
measurement. In \autoref{f:sbpdist}, we show the surface brightness 
distribution of the real galaxy 1D profiles measured at 1-4 \reff{}. 
The lower envelope of the projected 
axis ratio distribution clearly increases with increasing radius (and therefore
decreasing surface brightness). We do not find such a trend for the injected
disk population, which samples b/a uniformly, over the same range in surface brightness.

While we do not test for the impact of asymmetric features that are not captured
by a single \sersic{} model, this test verifies that
the ellipticity profile is well-recovered in HSC imaging conditions
across a range of different injected distributions.

\subsubsection{Comparison to parametric b/a measurements}
In order to capture the often irregular and asymmetric structure of dwarf 
galaxies, our non-parametric profile fits allow many parameters to 
vary as a function of semi-major axis. Thus, it is important to compare our
non-parametric measurements of ellipticity to measurements of the same quantity using
a more rigid model. 

In the left panel of \autoref{f:basanity}, 
we show the distribution over the observed axis ratio,
$q={\rm b/a}=1-\epsilon$, as measured from the non-parametric ellipticity profiles at \reff{} 
(filled teal histogram), from \sersic{} fits to the HSC-SSP imaging (green, unfilled),
and from exponential fits to SDSS imaging of the same galaxies. We find that 
our non-parametric measurements are in good agreement with both the 
\sersic{} profile fits ($\sigma_{\Delta q} = 0.07$) and the SDSS exponential
profile fits ($\sigma_{\Delta q} = 0.101$). The increase in scatter with respect to
the SDSS measurements is not unexpected, as the SDSS imaging is significantly shallower and
fit with a less flexible model (i.e. where the \sersic{} index is fixed to $n=1$). The lack of
bias as a function of projected axis ratio, however, is a good indication that our
non-parametric measurements obtain reasonable results despite their flexibility.

\begin{figure*}[htb]
\centering     
\includegraphics[width=.7\linewidth]{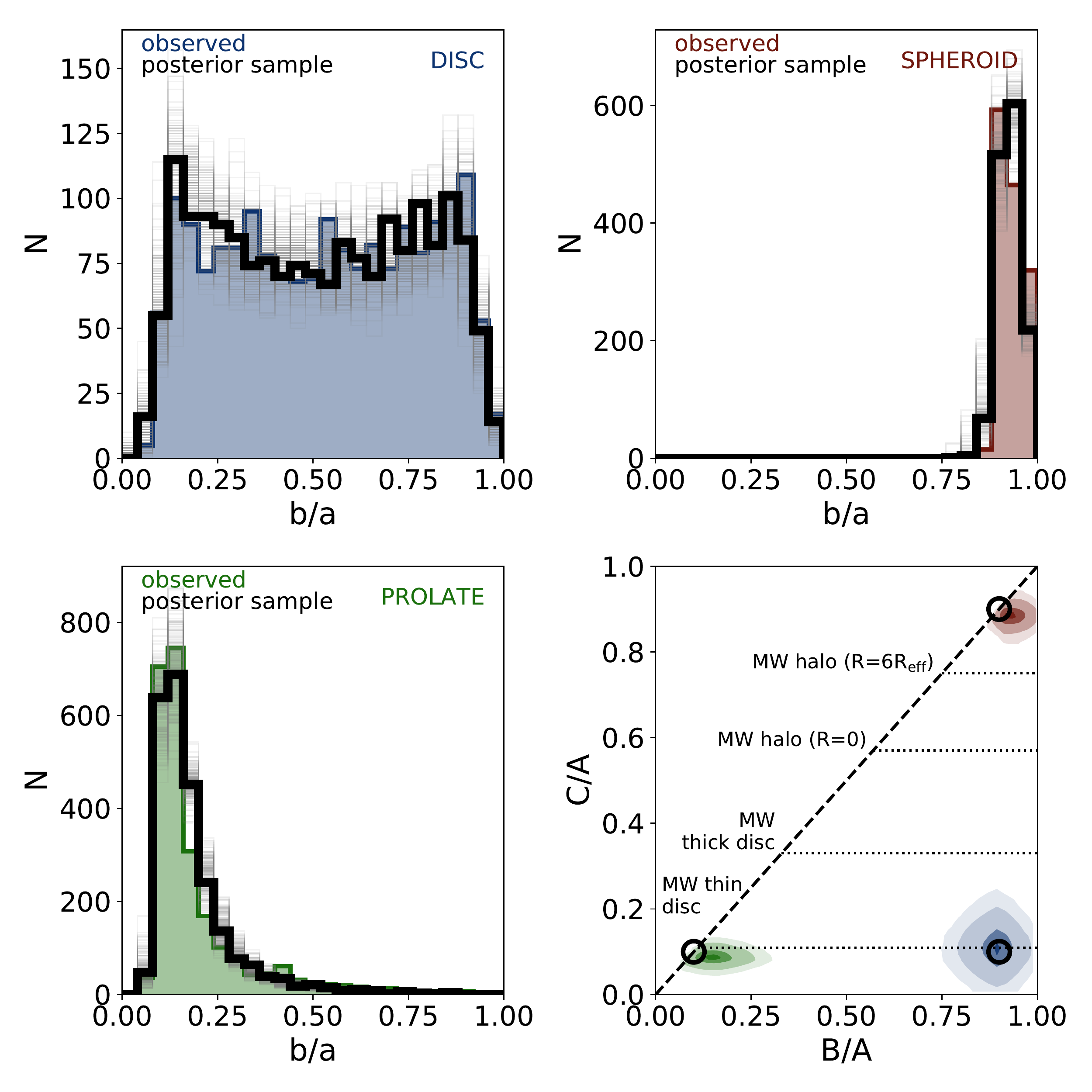} 
\caption{ 
    For the injected ellipsoid populations described in \autoref{f:extremerecovery}, we infer the 3D shape distribution
    of the population from the recovered projected axis ratios. 
    In each column, we show the inference results 
    for our injected populations of
    disk ($[A,B,C]=[1.,0.9,0.1]$, left), 
    spheroid ($[A,B,C]=[1.,0.9,0.9]$, middle), 
    and prolate ($[A,B,C]=[1.,0.1,0.1]$, right) galaxies. The top row
    shows the recovered b/a distribution as filled histograms and
    the posterior b/a sample as unfilled black histograms. The bottom row
    shows the distribution of the posterior sample over the intrinsic
    axes (B and C). The black circle shows the true position of the
    intrinsic axes. For physical context, we show the $C/A$ values measured
    for the Milky Way disk-halo system as measured by 
    \citet{schonrich2009} and \citet{iorio2019}. The dashed black line
    shows the definitional boundary of $B=A$.
    }
\label{f:boundaries}
\end{figure*}

\section{3D Shape Inference}\label{s:3dshapeinference}
Let us assume that the 3D shapes of each galaxy in our sample (or subsample) 
are drawn from a single distribution over the intrinsic 
axis ratios $B/A$ and $C/A$, 
given by P$(\vec \alpha)$ where $\vec\alpha$ is some set of parameters
that describes the distribution of $B/A$ and $C/A$.
The projected axis ratio, $q$, for any given ellipsoid is determined solely by the
observer's viewing angle, $(\theta, \phi)$. That is to say, the projected axis
ratio $q$ can be written as $q = \mathcal{F}(B/A,C/A,\theta,\phi)$.

The analytic expression for $\mathcal{F}$ was presented by \citet{simonneau1998},
and is reproduced below. First, $(ab)^2$ and $(a^2 + b^2)$ can be rewritten 
as follows:
\begin{equation}
\begin{split}
     a^2b^2=f^2&=(C\sin\theta\cos\phi)^2 +(BC\sin\theta\sin\phi)^2 +\\
     &(B\cos\theta)^2,\\
\end{split}    
\end{equation}

\begin{equation}
\begin{split}
    a^2+b^2=g&=\cos^2\phi + \cos^2\theta\sin^2\phi +\\
    & B^2(\sin^2\phi + \cos^2\theta\cos^2\phi) + (C\sin\theta)^2\\    
\end{split}    
\end{equation}
We now define the quantity $h$ to be
\begin{equation}
    h\equiv \sqrt{\frac{g-2f}{g+2f}},
\end{equation}
such that it may be shown that
\begin{equation}
    \frac{b}{a} = \frac{1-h}{1+h}
\end{equation}
Because the distribution of 
viewing angles is known to be isotropic on the surface of the sphere, 
we can predict the 
projected distribution of $q$ given a choice of intrinsic
shape distribution characterized by $\vec \alpha$ 
by sampling $\phi$ and $\theta$ as follows:

\begin{equation}
\begin{split}
    \phi &\sim \mathcal{U}[0,2\pi]\\
    \nu &\sim \mathcal{U}[0,1]\\
    \theta&=\cos^{-1}(2\nu - 1)\\
\end{split}
\end{equation}

For simplicity, we first
consider a normal distribution over both $B$ and $C$, such that the 
3D shape distribution can be described by the parameters 
$\vec \alpha = \{\mu_B,\mu_C,\sigma_B,\sigma_C\}$. We find that the 
data are well-described by this relatively simple model, and that the
fit is not significantly changed or improved by a more complex model  \citep[as 
motivated by][see \autoref{s:shapesize}]{zhang2019}.

Armed with this framework, we are able to quickly estimate the
distribution of projected axis ratios for a given choice of
$\vec \alpha$. This can then be compared cheaply to the observed distribution of 
$q$ by adopting a Poisson likelihood,
\begin{equation}
\rrr{\ln{}}p(q|\mu_B,\mu_C,\sigma_B,\sigma_C) = \sum_i n_i \ln{m_i} - m_i - \ln{n_i!},
\end{equation} 
where $n_i$ is the observed count where $0.04i<q\leq0.04(i+1)$ and $m_i$ is the
predicted count in the same range. Though this likelihood is in principle sensitive to
the adopted bin size, for our sample we find that the results are not
significantly affected by reasonable choices for the bin width. For each step, we
choose the binsize from a uniform distribution bounded by [0.03,0.1].
The minimum width is chosen such that for the minimum sample size
that we consider ($N=700$, see \autoref{s:massradius_3d}), 
for a uniform distribution of projected axis ratio the
standard deviation of the counts in a given bin is expected to
be $\sim20\%$ of the mean bin count.

We adopt a flat prior for all model parameters. The prior over $\mu_B$ and $\mu_C$ is
set purely by the physical boundaries:
\begin{equation}
p(\mu_B)=
\begin{cases}
1 \quad{\rm if}\quad{} 0<\mu_B<1 \\
0 \quad{\rm otherwise}
\end{cases}
\end{equation}
we additionally constrain $\mu_C\leq\mu_B$ to maintain the order of axes,
\begin{equation}
p(\mu_C)=
\begin{cases}
1 \quad{\rm if}\quad{} (0<\mu_C<1)&(\mu_C\leq\mu_B) \\
0 \quad{\rm otherwise},
\end{cases}
\end{equation}
additionally, when sampling from a given $\alpha$, we disregard cases where $C>B$.

We implement the same flat prior over $\sigma_B$ and $\sigma_C$:
\begin{equation}
p(\sigma_X)=
\begin{cases}
1 \quad{\rm if}\quad{} 0<\sigma_X<0.5 \\
0 \quad{\rm otherwise}
\end{cases}
\end{equation}
where $X \in{} (B,C)$. Here, the upper limit is set so that the distribution is
contained largely within the physically admissible region.\footnote{We furthermore find that 
none of our data suggest $\sigma_X$ near 0.5, indicating that this choice of boundary
does not affect our results.}

We can then write the posterior probability distribution as $p(\vec\alpha|q_{\rm obs})\propto p(q_{\rm obs}|\vec \alpha)p(\mu_B)p(\mu_C)p(\sigma_B)p(\sigma_C)$.
To sample efficiently
from this distribution, we use the Markov Chain Monte Carlo ensemble sampler implemented
in \textsf{emcee} \citep{foremanmackey2013}. For each case discussed in the work, we
run the sampler with 32 walkers and 3000 moves. We verify that the walkers have
converged and discard the first 150 moves of each.

\begin{figure*}[htb]
\centering     
\includegraphics[width=\linewidth]{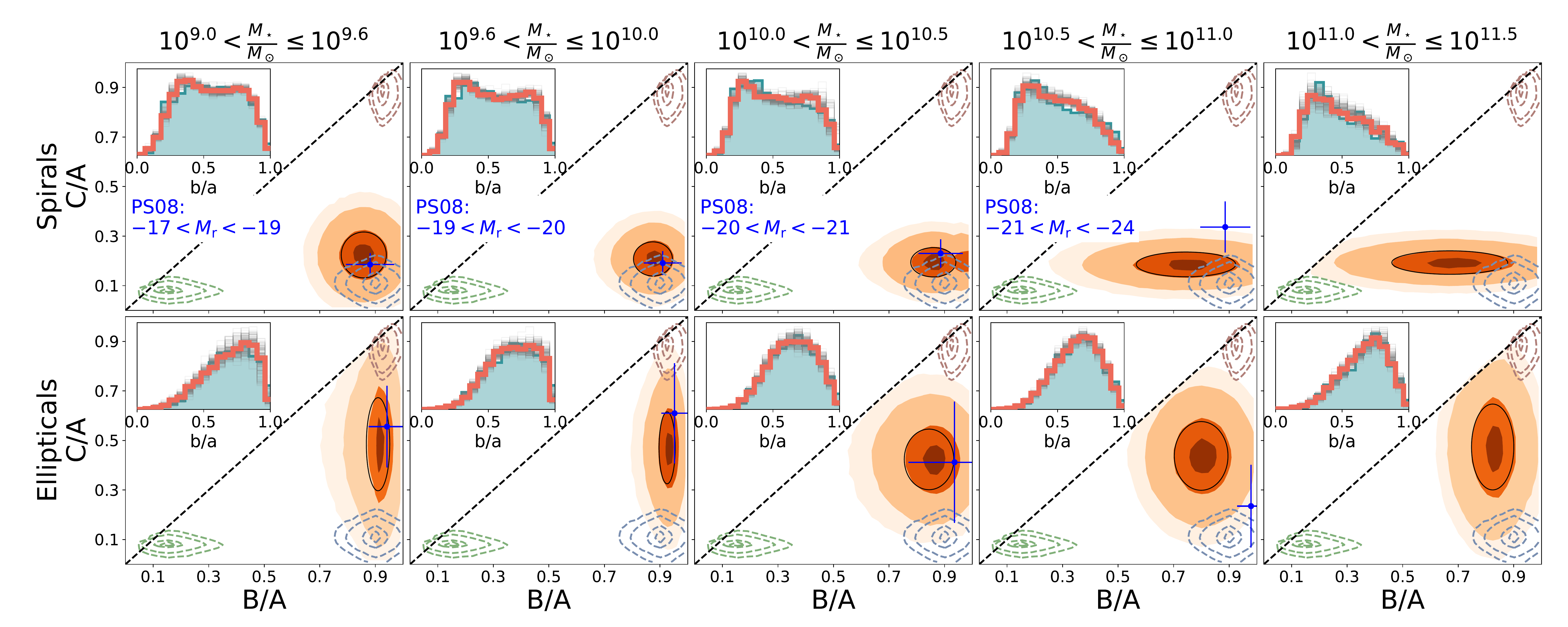} 
\caption{ 
    Inferred 3D shape distributions as a function of stellar
    mass (increasing towards top) and morphology (spirals at top,
    ellipticals at bottom), using SDSS catalog measurements following
    the morphology criterion of \citet{padilla2008}. In each panel, the
    filled contours show the inferred 3D shape distribution using our
    method, while the blue errorbars show the mean and standard deviation of
    the 3D shape distribution from \citet{padilla2008} in roughly analogous
    bins of $r_{\rm SDSS}$-band absolute magnitude. Our inferred shape 
    distributions are in general agreement with those of \citet{padilla2008},
    taking into account that the absolute magnitude cuts are not 
    equivalent to our cuts in stellar mass. In particular, the highest
    luminosity bin of \citet{padilla2008}, where our inference is in significant
    disagreement, extends to $M_\star \sim 10^{12} M_\odot$. 
    We also show the same contours for the pure disk, prolate, and spheroid \sersic{} population 
    presented in \autoref{f:boundaries} by the dashed blue, green, and red
    curves. 
    }
\label{f:padilla_intrinsic}
\end{figure*}

\begin{figure*}[htb]
\centering     
\includegraphics[width=\linewidth]{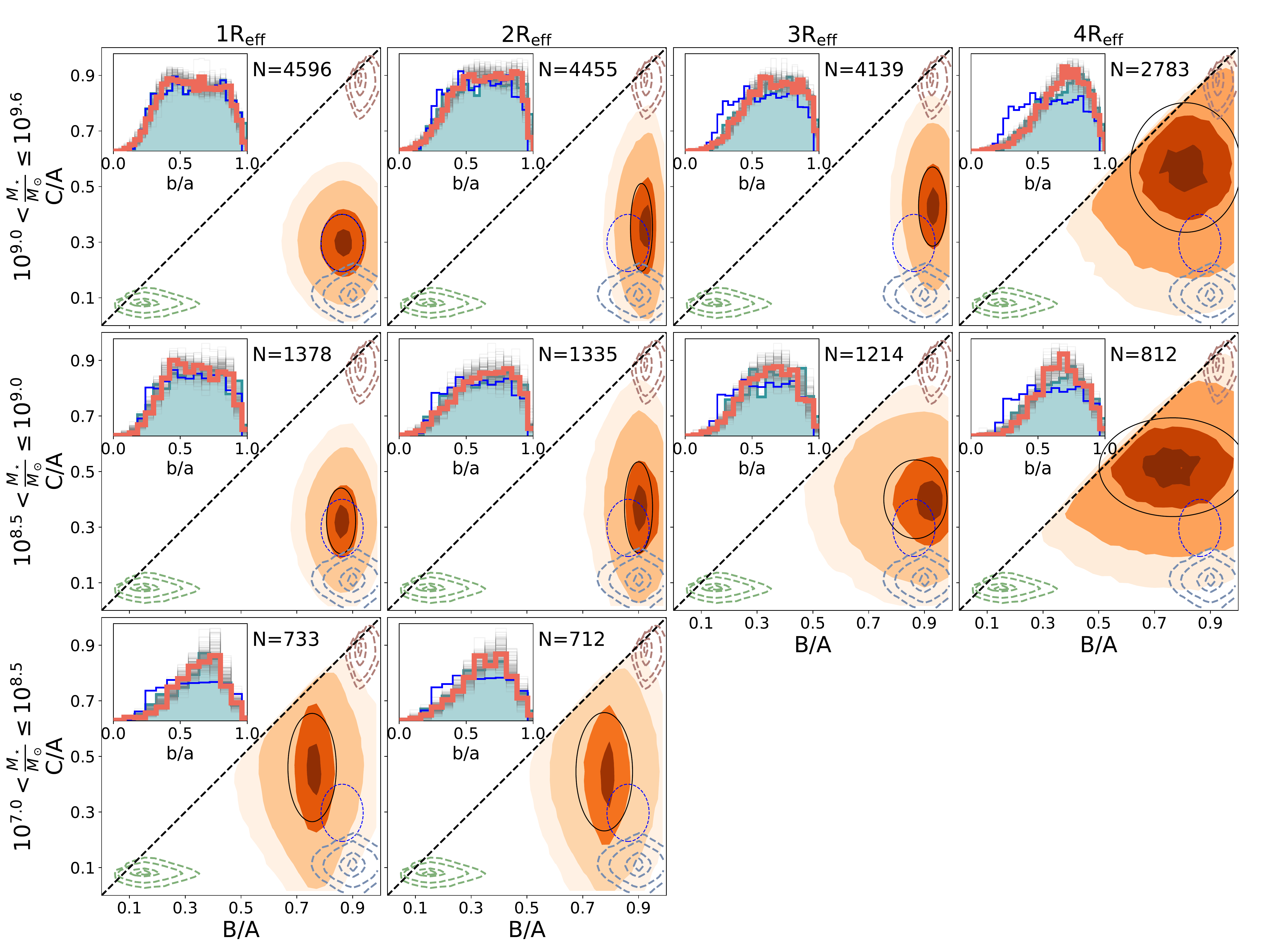} 
\caption{ 
    For bins in increasing stellar mass (rows) and measurement radius (columns),
    the distribution of inferred axis ratios B/A and C/A for the dwarf sample.
    Each panel shows the distribution of intrinsic axis ratios in orange. 
    The black ellipses show the
    1$\sigma$ region of the maximum a posteriori estimate, $\max_{P(\vec 
    \alpha|\{b/a\})} \vec \alpha$. 
    We also show the same contours for the pure disk, prolate, and spheroid \sersic{} population 
    presented in \autoref{f:boundaries} by the dashed blue, green, and red
    curves. The black line shows the B/A=C/A definitional boundary.
    Each inset panel shows the observed axis ratio, b/a$=1-\epsilon$, 
    distribution as the solid teal histogram. The distribution of b/a produced
    by the posterior sample in B/A and C/A is shown by the orange unfilled histogram;
    individual samples from the posterior are shown by the grey unfilled 
    histograms.
    }
\label{f:obsintr}
\end{figure*}

\subsection{Validation: Comparison to \sersic{} Populations}
In order to test our 3D inference framework, we return to the
injected \sersic{} populations of \autoref{s:sersicrecovery_2d}. These
tests have the advantage of incorporating both major sources of uncertainty in
the final 3D shape inference: the uncertainty in the 1D profile measurement (due to,
e.g., neighboring galaxies, residual sky background) and in the 3D shape inversion
problem. 

In \autoref{f:boundaries}, we show the recovered distribution of projected axis ratios by the
colored histograms: blue for the disk population, red for spheroidal, and green for
prolate. The distribution of projected axis ratios generated by sampling the posterior is
shown by the thick black stepped curve in each (the thin black lines show individual draws
from the posterior). The lower right panel of \autoref{f:boundaries} shows the distribution
of this posterior sample in the intrinsic B-C axis space; the colored contours show the
regions that contain $0.34^2$, $0.68^2$, $0.95^2$, and $0.99^2$ of the distribution (corresponding to the $0.5\sigma$,
1$\sigma$, 2$\sigma$, and 3$\sigma$ regions for a multivariate normal
distribution). The black circles show
the true value of $\mu_B$ and $\mu_C$.

When the true value of $\mu_B$ and $\mu_C$ are sufficiently distant from the boundary, we 
find that we are able to recover their values very well, as seen for the disk population in blue.
However, we find that our inferred values 
are biased when the true value is close to the 
imposed boundary (B/A$=$C/A) of the problem, as is the case for the spheroidal and prolate populations. Based on the shape distributions inferred by
studies at higher masses \citep[see,e.g.][]{padilla2008, vanderwel2014} and 
by studies of ultra diffuse galaxies in clusters \citep{burkert2017, 
rong2019}, as well as the shape distributions measured from
cosmological simulations \citep{pillepich2019}, it is unlikely that
real galaxies are characterized by extreme distributions
at the B$=$C boundary.
 
All of our injected \sersic{} galaxies are drawn from a 3D shape distribution
where $\sigma_B=0$ and $\sigma_C=0$; the inferred $\sigma_B$ and $\sigma_C$ in these
test cases should then provide a lower limit on the intrinsic dispersion to which we 
are sensitive.\footnote{Because the \sersic{} profiles are quite different from real
galaxies, we do not attempt to deconvolve the error in $\sigma_B$ and $\sigma_C$ for the
real sample using these test cases in this work.}

\subsection{Validation: Comparison to \citet{padilla2008}}\label{s:padillavalidation}
Before applying this shape inference framework to our sample of
HSC dwarfs, we want to confirm that our method can reproduce 
published 3D shape distributions of higher mass galaxies. Towards this end,
we use SDSS catalog shape measurements to infer 3D shape distributions
from data that are analogous to those of \citet{padilla2008}, who use SDSS catalog
measurements to estimate the 3D shape distribution of galaxies with 
stellar masses of $M_\star \gtrsim 10^9 M_\odot$. 

For a linear combination
of an exponential profile (i.e., a \sersic{} profile with index $n=1$) and a
deVaucouleurs profiles (i.e., a \sersic{} profile with index $n=4$), the SDSS
photometric catalog provides the weight assigned to the $n=4$ component
as $f_{\rm deV}$ \citep{sdssdr2}.
\citet{padilla2008} separate their sample into a subset of spiral galaxies,
wherein $f_{\rm deV}<0.8$, and elliptical galaxies, wherein $f_{\rm deV}>0.8$, and 
evaluate their 3D shape distributions independently. Using the SDSS DR16 
catalog, we divide the sample of galaxies with $z<0.05$ and
$M_\star \gtrsim 10^9 M_\odot$ in the same manner \citep{sdssdr16}. 

We then use the 3D shape inference method described above to estimate the
distribution over intrinsic axes B and C for bins of approximately 0.5 dex in stellar mass
(the first mass cut is set at 
$\log_{10}(M_\star/M_\odot)=9.6$ so that the lowest mass bin considered is
equivalent to the highest mass bin of 
our sample).

The results of this inference, with the roughly equivalent $r_{\rm SDSS}$-band
absolute magnitude bins of \citet{padilla2008} overlaid, are shown in
\autoref{f:padilla_intrinsic}. 
Each panel inset shows both the observed distribution of b/a, measured via the 
non-parametric profile construction described in \autoref{s:1dprofiles},
by filled teal histograms. We then overplot the distribution of projected
b/a generated from sampling the MCMC chain in grey, i.e. $\vec\alpha_i \sim P(\mu_B,\mu_C,\sigma_B,\sigma_C|{\rm b/a})$. The distribution of b/a over all
of these samples is then shown by the thick orange curve. In the main
panel, we show the 
analogous distribution in the intrinsic axes (B-C) plane. 
Here, the same samples from the posterior are
shown by the orange filled contour. The dashed colored curves are the 
contours of the pure disk, spheroid, and prolate \sersic{} populations shown in
the lower right panel of \autoref{f:boundaries}.

Due to the differences in data used and the
imperfect mapping between stellar mass and absolute magnitude, we do not
expect that the inferences will be statistically identical.
However, we find good agreement between our results and 
those of \citet{padilla2008} for all but their brightest bin 
at $-21<M_r<-24$, which includes a significantly wider mass range 
than our most similar bin. Having reproduced the 3D shape 
distribution of this literature sample, and of the mock galaxies
injected into HSC imaging, we proceed to apply the shape 
inference technique to the dwarf sample at hand.

\begin{figure*}[htb]
\centering     
\includegraphics[width=\linewidth]{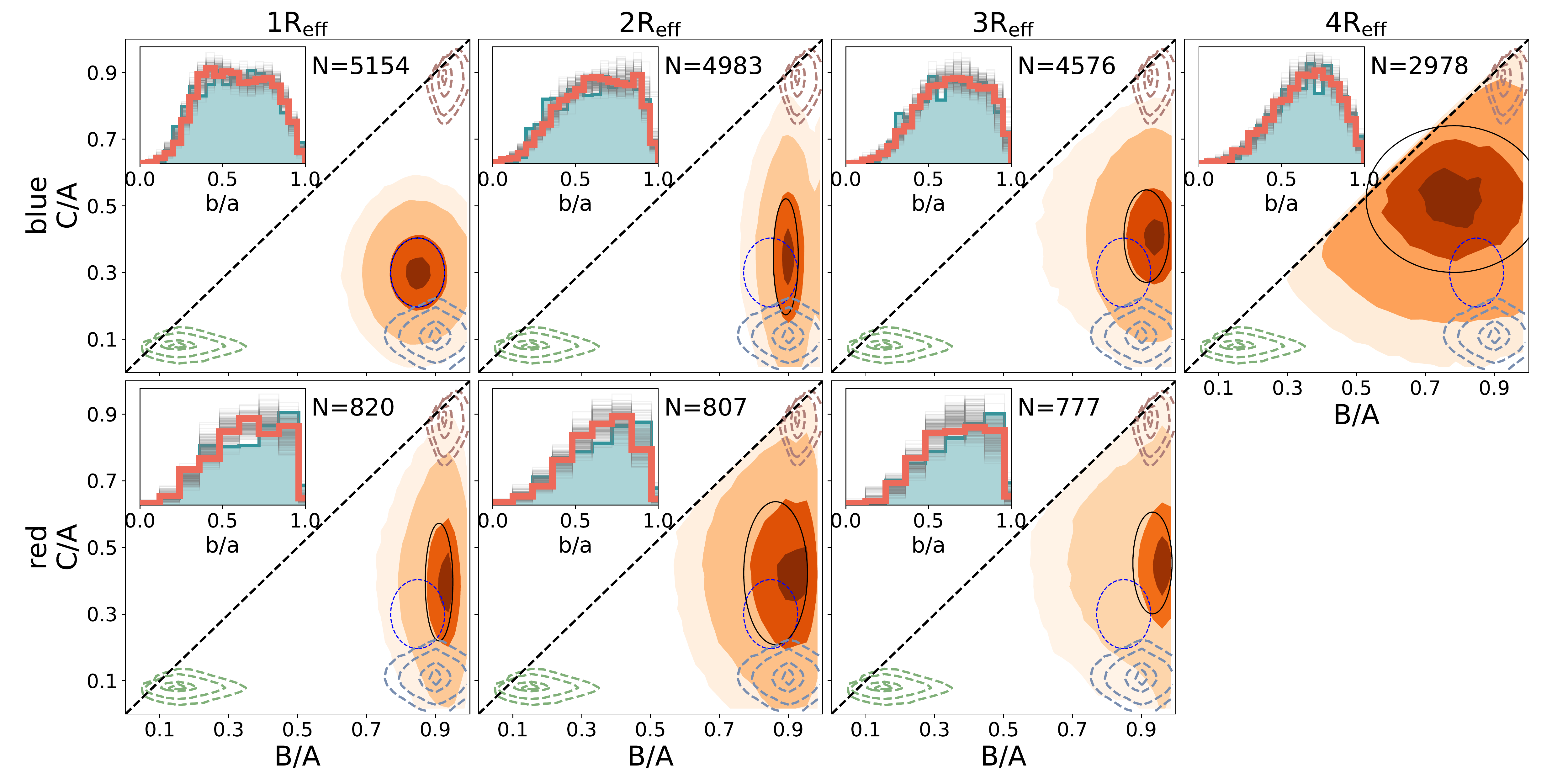} 
\caption{ 
    For dwarfs with $M_\star>10^{8.5}M_\odot$, the inferred shape 
    distribution for blue ($g-i<0.90$) and red ($g-i>0.90$) galaxies as a 
    function of radius. Similar to the trend we see in the analogous mass 
    bin of \autoref{f:padilla_intrinsic}, the 3D distribution of blue galaxies
    is diskier than the red galaxies at $R=1$\reff{}. At large radii, both
    blue and red galaxies move towards spheroid shapes. 
    As in \autoref{f:obsintr}, we show the 50\thh{} percentile 
    regions of the pure disk, prolate, and spheroid \sersic{} population 
    presented in \autoref{f:boundaries} by the dashed blue, green, and red
    curves.
    }
\label{f:colorsep}
\end{figure*}

{}

\begin{figure*}[htb]
\centering     
\includegraphics[width=\linewidth]{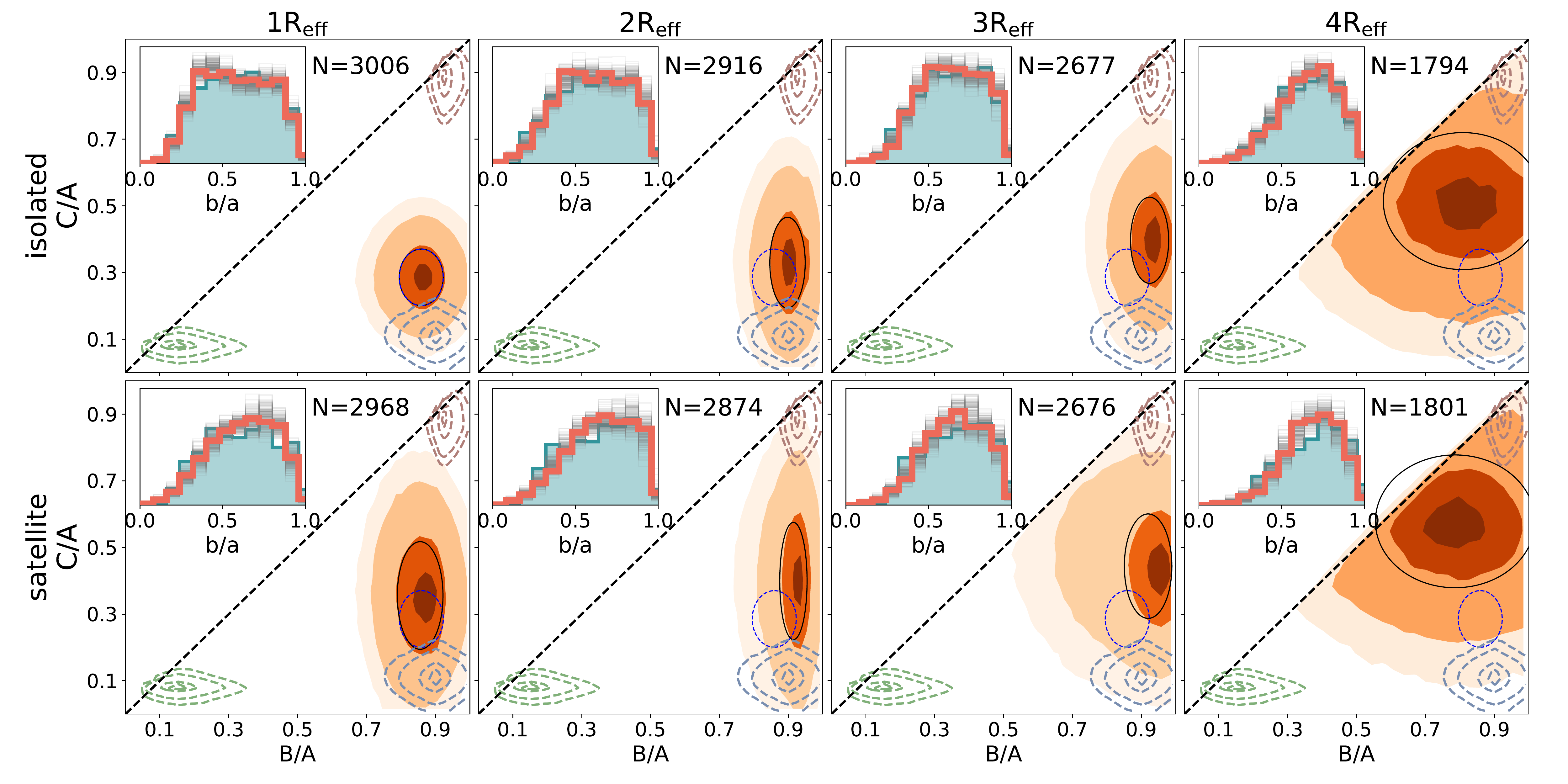} 
\caption{ 
    For dwarfs with stellar masses $M_\star>10^{8.5}M_\odot$, the inferred shape 
    distribution for field (distance to nearest neighbor ${\rm d_{NN}} > 1$Mpc) and satellite (${\rm d_{NN}} \leq 1$Mpc) galaxies. We see 
    negligible difference between satellite and field galaxies; both 
    are consistent with thick discs at 1\reff{}, and C/A increases towards 
    spheroid shapes at larger distances from the galaxy center. 
    As in \autoref{f:obsintr}, we show the analogous contours
    of the pure disk, prolate, and spheroid \sersic{} population 
    presented in \autoref{f:boundaries} by the dashed blue, green, and red
    curves.
    }
\label{f:dnnsep}
\end{figure*}

\section{Results}\label{s:results}
The HSC-SSP imaging has both the surface brightness 
sensitivity to reliably measure the ellipticity profile of individual 
galaxies out to 3\reff{} and the on-sky area necessary to build a 
sample large enough to infer the 3D shape distribution. 

Before proceeding, we note that the 
transition from thin disk to thick disk to spheroid does not have a well-defined
boundary. To put our results into context, we
provide the intrinsic axis ratios of the Milky Way disk and halo
system.
Assuming an 
exponentially declining disk, 
the Milky Way has a disk scale length of 
2.5 kpc, a thin disk scale height of 270 pc, and a thick disk
scale height of 820 pc. This corresponds 
to an axis ratio of $C/A=0.11$
for the MW thin disk, and $C/A=0.33$ 
for the MW thick disk \citep{schonrich2009}. 
The MW stellar halo becomes increasing
spheroidal as a function of radius in the 
range of $0.57\leq C/A \leq 0.75$ \citep{iorio2019}.

\subsection{3D Shape as a function of Stellar Mass and Radius}\label{s:massradius_3d}
We first consider the change in the galaxy 3D shape as a function of
stellar mass and radius. Previous studies have found that the high mass
end of our sample ($10^9 < M_\star/M_\odot <10^{9.6}$) are composed 
largely of (thick) discs \citep{padilla2008, sanchezjanssen2010, vanderwel2014}; 3D shapes beyond
1\reff{} or at lower masses have not been measured for general dwarf samples.

First, we consider the distribution of observed b/a as measured at 1\reff{}, 
2\reff{}, 3\reff{}, and 4\reff{} for bins of stellar mass in \autoref{f:obsintr}.
Each panel shows the number of measurements made for each slice in
radius and stellar mass. 
From tests with \sersic{} populations, we find that the recovered 
values of $\sigma_B$ and $\sigma_C$ increase significantly at $N\lesssim700$;
we therefore do not consider subsets where $N<700$ (see \autoref{s:samplesize}).

From the observed distribution alone, we see first that the 
distribution of projected axis ratios
becomes increasingly more concentrated at large values of b/a as we consider
larger radii. This is consistent with a shift towards more spheroidal shapes;
indeed, in \autoref{f:obsintr}, we see 
that the high-mass dwarfs are consistent with a thick disk at 1\reff{}, 
and shift towards the spheroidal extreme at 3\reff{} and 4\reff{}. We note that
a minority of the galaxies in the sample have bars at 1\reff{} -- though these bars
clearly do not dominate the signal (bars are intrinsically prolate, with 
$\mu_B\sim\mu_C\sim0.3$ \citet{compere2014,mendezabreu2018}), it is likely that
they contribute to the recovered triaxiality of the sample. Indeed, though the
dwarfs are only slightly triaxial ($\mu_B \sim 0.75$ at all masses and radii),
we find that a purely oblate model is a significantly worse fit to the data. 

We see relatively little change in the intrinsic shape distribution of galaxies
at $10^{8.5} < M_\star/M_\odot \leq 10^{9.0}$ and those at
$10^{9.0} < M_\star/M_\odot \leq 10^{9.6}$. 
Both show $\mu_C(1{\rm R_{eff}})\sim 0.3$, and shift towards progressively larger
$\mu_C$ with increasing radius. In our lowest mass bin, we see a hint that the 
shape distribution shifts dramatically, towards lower $\mu_B$ and $\mu_C$ (i.e.
away from the pure disk region). Though this behavior is not unexpected, as
lower mass galaxies are expected to be increasingly dispersion dominated
\citep[see, e.g.][]{wheeler2017,pillepich2019}, 
we caution that we are not complete at this mass bin, 
and that the number of galaxies in our lowest mass bin is significantly
lower (N$\sim700$), which may lead to an overestimation of $\sigma_B$ 
and/or $\sigma_C$ (see \autoref{s:samplesize}).

\subsection{3D Shape and Galaxy Color}\label{s:colorsep}
At higher masses, we reproduce in \autoref{f:padilla_intrinsic}
the divergence in 3D shape distribution of spiral and elliptical galaxies 
seen in \citet{padilla2008}. One can then reasonably expect to see a similar
trend at $\approx 1$\reff{} when our sample is divided between red and 
blue galaxies. We divide our sample at $(g-i)_{\rm SDSS}=0.9$, 
chosen as the 
midpoint between the blue sequence and red cloud for this choice of color
and stellar mass range\footnote{Because the red sequence is relatively unpopulated at this mass range, we choose the division using the SDSS catalog and a somewhat broader range in stellar mass $M_\star \lesssim 10^{10.5}M_\odot$}. 

We show the 3D shape distribution for the highest mass bin in \autoref{f:colorsep}
as a function of radius, again at 1\reff{}, 2\reff{}, and 3\reff{} from left to right.
The left column shows the results at 1\reff{}. Indeed, in the leftmost column
we see that the red galaxies are at preferentially
larger C/A with respect to blue galaxies, similar to 
\autoref{f:padilla_intrinsic} when the sample is split between spiral and 
elliptical galaxies. 

At larger radii, however, we find that the blue and red galaxies occupy the same region 
in the B-C plane. 
While the red galaxies show little shape change with radius, 
the distribution of blue galaxies increases in 
$\sigma_C$ and shifts towards the spheroidal 
corner (red dashed curve) of parameter space.

\subsection{3D Shape and Environment}\label{s:dnnsep}
Star formation in dwarfs is thought to be quenched by almost entirely
environmental means \citep{geha2012}; it is of interest, then to 
ask whether populations of field and satellite dwarfs display the same
change in 3D shapes as do blue and red dwarfs. For this exercise, 
we search massive companions ($M_\star > 10^{10}M_\odot$) in the NASA
Sloan Atlas (NSA)
within $\Delta v < 1000 \kms{}$ and 1 Mpc projected distance. 
We choose the 
projected distance cut $d_{\rm NN} > 1$Mpc to coincide with 
the distance at which the 
dwarf quenched fraction, $f_{\rm quench}(d_{\rm NN})$, approaches its field limit
$f_{\rm quench}(d_{\rm NN} \rightarrow \infty)$
 for the mass range considered in \citet[][see Figure 4]{geha2012}.
We additionally consider only galaxies at $z<0.10$. Though
the NSA contains galaxies up to $z=0.15$, at $z>0.1$ the satellite fraction
begins to drop, indicating that the massive galaxy sample is not sufficiently
complete to characterize the environment of the dwarfs in our sample.
In \autoref{f:dnnsep}, we show the 3D shape distribution for
field (top row) and satellite (bottom row) dwarfs at $M_\star>10^9M_\odot$. 
Unlike the dwarfs separated by color, the satellite and field
dwarf samples show roughly the same 3D shape distribution as a function of
radius. It is important to note that the difference in $\sigma_B$
at large radius between the field and satellite dwarfs is likely unphysical,
as the inferred $\sigma_B$ for these populations is close to or below the 
estimated $\sigma_B$ for our zero-scatter disk and spheroid populations (shown
in \autoref{f:dnnsep} by the dashed contours). 

We find that at 1\reff{}, the satellite galaxies scatter towards
marginally more spheroidal shapes than do the field galaxies. This 
effect is similar to the trend seen in \autoref{s:colorsep} between red
and blue dwarfs, but the
separation between field and satellite galaxies is relatively small. 
Though the projected axis ratio
distribution of field galaxies and satellite galaxies is significantly
different (Kolmogorov-Smirnov p-value$\approx0.001$), we do not 
see a significant difference between the $b/a$ distribution of
field galaxies and blue satellites (KS p-value$\approx0.29$). 
This is likely due to fact that, for our sample, nearly all red galaxies are 
satellites of more massive galaxies \citep{geha2012}. Blue galaxies, on the
other hand, are found both as satellites and in the field.
It is likely that making a cut on color produces, 
in effect, a selection of satellite galaxies
that have been more processed by the host central. 

\begin{figure}[htb]
\centering     
\includegraphics[width=\linewidth]{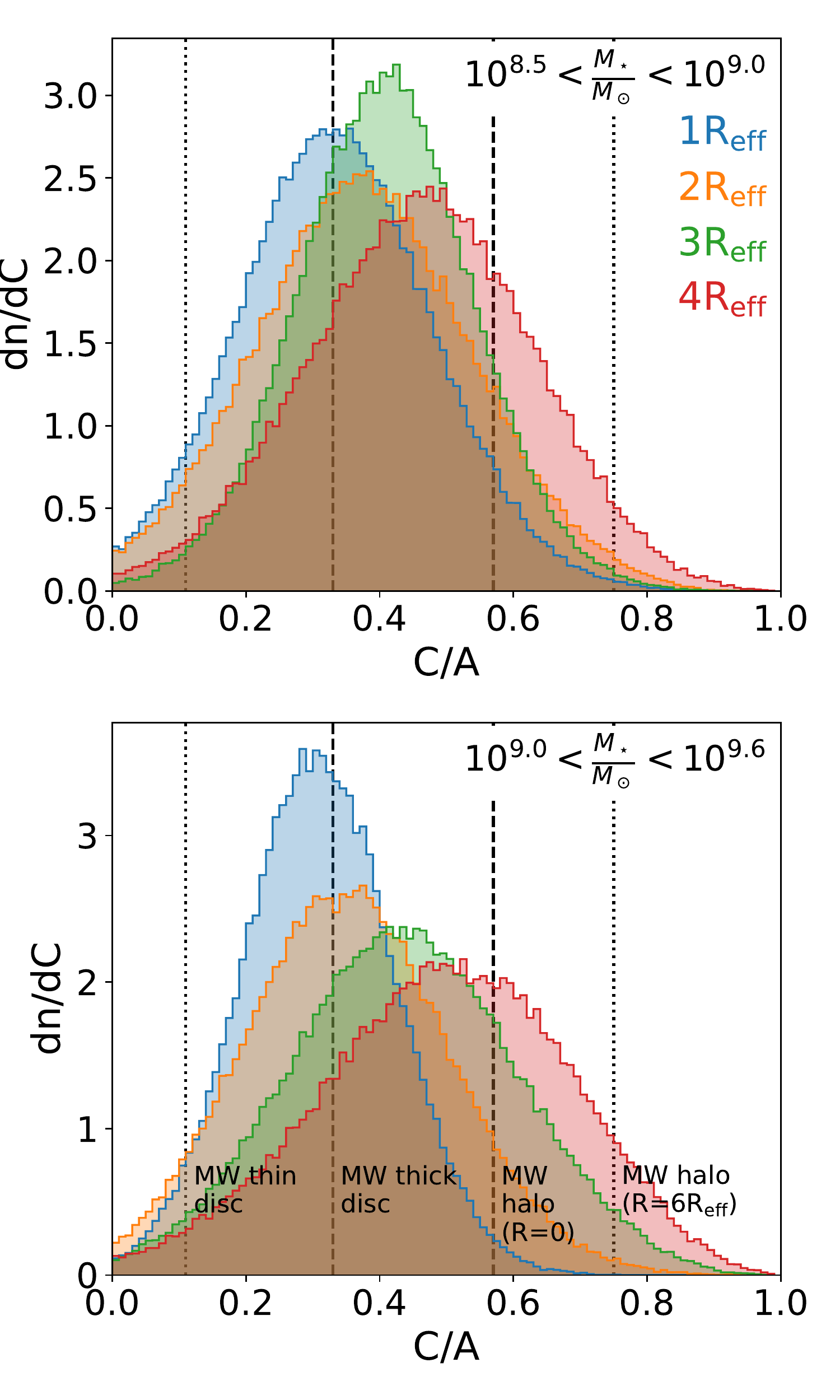} 
\caption{ 
    The change in galaxy thickness (C/A) as a function of radius for
    $10^9 < M_\star/M_\odot < 10^{9.6}$ (top) and 
    $10^{8.5} < M_\star/M_\odot < 10^{9}$ (bottom). In both cases, the
    dwarfs become systematically more spheroidal (C$\sim$A, $B/A\gtrsim0.8$ for
    all cases) at large 
    radii. For physical context, we also show the
    $C/A$ axis ratio of the Milky Way thin disk \& thick disk
    \citep{schonrich2009}, and the
    stellar halo at $R=0$ \& $R=6$\reff{} \citep{iorio2019}.
    The MW axis ratios are shown by
    vertical black lines, as labeled in the bottom panel.
    }
\label{f:cachange}
\end{figure}

\section{Discussion}\label{s:discussion}

\subsection{The Emergence of Round Outskirts \rrr{A}round Low-Mass Galaxies}
At higher masses, stellar halos are 
generally thought
to be the product of a series of minor mergers, which deposit stars at large radii \citep{amorisco2017}. 
However, because the stellar-to-halo mass ratio 
tends to increase as galaxy mass decreases,
the dwarfs in our sample are unlikely to 
accrete a sufficient mass in stars to build a stellar halo via 
minor mergers alone \citep{purcell2007,moster2013}. 

Even though it appears unlikely that dwarfs can accrete a stellar halo
via conventional means, 
it has long been known that dwarfs in the Local Group and M81 Group
host a smooth 
intermediate/old stellar population in their outskirts \citep[see ][and references therein]{stinson2009, hargis2020}. These observations hint at the existence of
an in-situ halo formation mechanism at low masses, but this conclusion is 
obfuscated for two reasons:, first the sample is comprised almost entirely of
galaxies that are interacting with a more massive companion.
Second, such stellar halos have been confirmed for only a few tens of galaxies. 
In this work, we have presented the 
first large sample where a clear 
transition to a round stellar component is detected in the 
outskirts of dwarfs that are not Local Group members. 

In \autoref{s:massradius_3d}, we presented a set of inferred 3D shapes
for a sample of dwarf galaxies as a function of stellar mass and radius.
Due to the depth of the HSC-SSP imaging, we are able to measure ellipticity
profiles out to 4\reff{} for the most massive dwarfs in our sample. Indeed,
we see that the structure of the dwarfs are characterized by thick discs at
1\reff{} and become increasingly spheroidal at large radii, as
shown in the top row of \autoref{f:obsintr}. For 
clarity we also show the C/A distribution as a function of 
radius for the $10^9 M_\odot<M_\star<10^{9.6}M_\odot$ and 
$10^{8.5} M_\odot<M_\star<10^{9}M_\odot$ stellar mass bins in 
\autoref{f:cachange}. At 1\reff{}, the dwarfs have $C/A$ axis
ratios consistent with the Milky Way thick disk, and 
significantly thicker than the MW thin disk \citep{schonrich2009}.
This thick disk morphology is in rough agreement with previous
measurements for dwarfs in SDSS and the Local Group 
\citep{padilla2008, sanchezjanssen2010, roychowdhury2013}.
At 4\reff{}, the dwarfs shapes are of comparable
thickness to the inner stellar halo of the MW as measured
by \citet{iorio2019}. 
These results 
indicate that the
dwarf disk-halo interface is similar in structure, if not origin,
to more massive galaxies.

Though they are not expected to form through minor mergers, 
the existence of
round stellar outskirts around dwarfs is not unexpected theoretically.
Dwarf galaxies
sit in shallow potential wells, and are thus more sensitive to the effects
of star formation feedback than their more massive analogs.
\citet{stinson2009} proposed that dwarf galaxies would generically
form stellar halos through stellar radial migration, star formation
in outflows, and a contraction of the central star-forming region.
Similarly, \citet{maxwell2012} found that stellar feedback could
drive sufficient quantities of dense gas to produce a fluctuation 
in the overall potential and thus build a stellar spheroid through
migration \citep[see also][]{elbadry2016}.
The concordance in 3D shape at large radii for 
blue and red dwarfs, as shown in \autoref{f:colorsep}, and at all 
measured radii in field and satellite dwarfs, as 
shown in \autoref{f:dnnsep}, 
also suggests that the creation of round outskirts
is not driven by an interaction with a more massive halo.

It has also been suggested that dwarf halos could be formed as a result
of major mergers between dwarfs \citep{bekki2008}. However, the detection of
an increasingly spheroidal component in the outskirts of our dwarf sample 
is at odds with this formation mechanism; the major
merger rate of dwarfs is likely not high enough to generate enough
stellar halos to produce such a population. Simulations
suggest that approximately 30\% of dwarfs in our stellar
mass range outside of the Virial radius of a MW-like object
have undergone a dwarf-dwarf major merger in the past 10 Gyr
\citep{deason2014}. Though dwarf-dwarf major mergers were more
common in the early universe, due to fading via passive evolution
it is unlikely that we would be able to detect such ancient halos.
\rrr{Though the $z=0$ surface brightness of any given 
merger-driven stellar halo is dependent on its assembly history,
\citet{bekki2008} finds that their simulated stellar halo reaches $\sim30$ mag 
arcsec$^{-2}$ at $R\sim 2$ kpc. If the stellar halo population had
a uniform age of 1 Gyr at the time of halo creation, and was created
at a lookback time of 10 Gyr, the stellar halo will have dimmed by 1-2
mag arcsec$^{-2}$ by $z=0$ from the evolution of the mass-to-light ratio
alone (as computed from the FSPS models of \citealt{conroy2009}), well below
the surface brightness sensitivity of our imaging.} 
Moreover, the intermediate age stellar component 
often observed at large
radii in resolved star studies requires a relatively
recent deposition of stars in the outskirts \citep[see, e.g.][and citations therein]{zaritsky2000,stinson2009}.
We thus find it unlikely that major mergers are the sole 
formation mechanism of low-mass stellar halo formation, though
we note that some individual cases are consistent with both
star formation feedback and accretion driving stellar halo 
formation \citep{pucha2019}. 
The apparent ubiquity of round stellar outskirts in this work is 
instead consistent with the proposal that stellar outskirts
are formed primarily through in-situ processes.

\subsection{Morphological Transformation and Quenching}
It has long been observed that the cessation of star formation in dwarfs, their morphological transformation from a disk-dominated
to a dispersion-dominated structure, and their proximity to more massive galaxies 
are all strongly correlated \citep[see, for example, ][]{dressler1980,lin1983,postman1984,weinmann2006,geha2012,kormendy2012,ann2017}. 

As shown in the leftmost column of \autoref{f:colorsep}, we find that
the blue dwarfs in our sample tend towards lower values of
$C/A$ than red dwarfs -- that is, the blue dwarfs are more consistent
with a thick disk, while red dwarfs scatter towards more 
spheroidal shapes. This is in concordance with the 
familiar morphology-color dichotomy, and follows the same
trend seen in intrinsic shape studies at higher masses 
\citep[see, e.g.][and \autoref{f:padilla_intrinsic}]{padilla2008,rodriguez2016}. We see
a similar trend when separating dwarfs by the projected 
nearest neighbor distance (within 1000$\kms{}$), though the shift
between satellite and field galaxies is relatively marginal
(see leftmost column of \autoref{f:dnnsep}). 

We do not detect a significant difference in the
intrinsic shapes of blue satellites and dwarfs in the field. 
This is in contrast with the shift towards rounder shapes seen 
in the red galaxy subset (\autoref{f:colorsep}) and 
satellite galaxies (without a color cut, \autoref{f:dnnsep}). 
This suggests that morphological transformation operates
on a longer timescale than star formation quenching, or that 
quenching is a prerequisite to the morphological transformation. 
However, it is important to note our choice of model (a singly-peaked
multivariate Gaussian) will necessarily only recover the dominant shape
population.



\subsection{Comparison with Simulations}
Due again to their increased sensitivity to star formation feedback,
the 3D shapes of dwarfs are a strong constraint on the feedback prescription
of cosmological simulations. 

\citet{pillepich2019} give the distribution of dwarf 3D shapes as measured
at twice the stellar half-mass radius. First, we note that our results
are in broad agreement with those of \citet{pillepich2019} in that our 3D
shapes are in the disky/spheroidal regime, with essentially no galaxies
in the prolate regime (defined by \citet{vanderwel2014} as $B/A<1-C/A$).
At \reff{}, the change in our 3D shape distribution is also qualitatively
similar to that of \citet{pillepich2019}; as stellar mass decreases, $\sigma_C$
increases and the distribution shifts towards more spheroidal 
shapes (as shown in the leftmost column of \autoref{f:obsintr}). 

However, under the assumption of centrally concentrated star formation,
we would expect the half-mass radius to be larger than the half-light radius,
implying that a comparison at the same physical radius should occur at
$>2$\reff{}. Moreover, the conversion between the $i_{\rm HSC}$-band
surface brightness and the stellar mass distribution is a function of
the galaxy's stellar populations. 

Clearly, to make a quantitative comparison, it will require significant
effort to put the simulations and observations on equal footing. Nevertheless,
the broad agreement in the shape distribution of dwarfs and its evolution
with stellar mass is a promising step.

\section{Conclusions}
In this work we have measured the surface brightness and 
ellipticity profiles of a sample of 
spectroscopically confirmed dwarfs using imaging from the Hyper Suprime-Cam
Subaru Strategic Program (\autoref{s:1dprofiles}). We then extended the
framework commonly used to infer 3D galaxy shapes \citep[see, e.g.,][]{padilla2008,roychowdhury2013,vanderwel2014,zhang2019, putko2019}
to measure the change in dwarf galaxy shape as a function of radius.

We show that the population of dwarfs in our sample tend to host
thick, disk-like structures at 1\reff{}, and evolve towards more
spheroidal shapes in their outskirts (see \autoref{f:obsintr}).
This finding is in agreement with the predicted 
quasi-spherical shapes of in-situ stellar halos \citep{stinson2009}. 

At $M_\star>10^{8.5}M_\odot$, blue dwarfs tend to be diskier than red 
dwarfs near their centers (i.e. at R=1\reff{}, left panel of 
\autoref{f:colorsep}). This divergence as a function of color mirrors
the same dichotomy seen at higher masses, where blue galaxies are 
preferentially diskier and red galaxies relatively thicker
and spheroidal (see \autoref{f:padilla_intrinsic}). 
However, the outskirts of both red and blue dwarfs move towards 
more spheroidal shapes, suggestive of an in-situ formation mechanism
for the extended stellar outskirts. This interpretation is
also supported by a uniform trend towards more spheroidal outskirts
in both field and satellite galaxies (see \autoref{f:dnnsep}).

The sample considered in this work is based on spectroscopic surveys;
we are thus missing low surface brightness and ultra diffuse galaxies (UDGs). In particular, 
simulations suggest that the effective surface brightness cuts implemented
in the SDSS and GAMA spectroscopic surveys bias dwarf samples towards
more compact objects \citep{wright2020}.
Previous works have focused on samples of cluster UDGs: 
\citet{burkert2017} find
that, for a sample of Coma UDGs, $\mu_B=\mu_C\sim0.67$ (for a model 
with the assumption $C=B\leq A$). Similarly, \citet{rong2019} find that for
a triaxial model, $\mu_B=0.86$ and $\mu_C=0.49$. These results imply that
cluster UDGs are typically rounder than high surface brightness dwarfs, with
an intrinsic minor axis ratio ($C/A$) comparable to the outskirts of 
dwarfs in our sample. 

However, samples of cluster UDGs are in highly overdense environments
compared to the typical dwarf galaxy in our sample. To more fairly compare the
structural composition of low surface brightness  
and high surface brightness dwarfs,
and better understand the nature of the relationship between these two
populations, we must instead look to build a sufficiently
large sample of UDGs in the field 
\citep[e.g., ][]{bellazzini2017,leisman2017,roman2017,greco2018,tanoglidis2020}
such that the deprojection problem is tractable. 

Our analysis also suggests that at $M_\star<10^{8.5}M_\odot$, dwarfs become
increasingly round (larger $\mu_C$). However, our spectroscopic sample
is far from mass-complete at $M_\star<10^8 M_\odot$. 
In order to more comprehensively understand the properties of the
dwarf population,
it is necessary to construct a large and mass-complete sample of dwarfs
at stellar masses lower than what is accessible with spectroscopic 
surveys currently in hand. As has been shown theoretically, 
star formation feedback is expected to play an increasingly
dramatic role in the stellar structure of increasingly low
mass dwarfs; extending the observational
lever arm to lower stellar masses will provide a important 
constraint on prescriptions for star formation feedback, and provide
novel insights into the stellar structure of such systems.

\acknowledgements
The Hyper Suprime-Cam (HSC) collaboration includes the astronomical communities of Japan and Taiwan, and Princeton University.  The HSC instrumentation and software were developed by the National Astronomical Observatory of Japan (NAOJ), the Kavli Institute for the Physics and Mathematics of the Universe (Kavli IPMU), the University of Tokyo, the High Energy Accelerator Research Organization (KEK), the Academia Sinica Institute for Astronomy and Astrophysics in Taiwan (ASIAA), and Princeton University.  Funding was contributed by the FIRST program from the Japanese Cabinet Office, the Ministry of Education, Culture, Sports, Science and Technology (MEXT), the Japan Society for the Promotion of Science (JSPS), Japan Science and Technology Agency  (JST), the Toray Science  Foundation, NAOJ, Kavli IPMU, KEK, ASIAA, and Princeton University.

This paper makes use of software developed for the Large Synoptic Survey Telescope. We thank the LSST Project for making their code available as free software at  http://dm.lsst.org

This paper is based [in part] on data collected at the Subaru Telescope and retrieved from the HSC data archive system, which is operated by Subaru Telescope and Astronomy Data Center (ADC) at NAOJ. Data analysis was in part carried out with the cooperation of Center for Computational Astrophysics (CfCA), NAOJ.

The Pan-STARRS1 Surveys (PS1) and the PS1 public science archive have been made possible through contributions by the Institute for Astronomy, the University of Hawaii, the Pan-STARRS Project Office, the Max Planck Society and its participating institutes, the Max Planck Institute for Astronomy, Heidelberg, and the Max Planck Institute for Extraterrestrial Physics, Garching, The Johns Hopkins University, Durham University, the University of Edinburgh, the Queen’s University Belfast, the Harvard-Smithsonian Center for Astrophysics, the Las Cumbres Observatory Global Telescope Network Incorporated, the National Central University of Taiwan, the Space Telescope Science Institute, the National Aeronautics and Space Administration under grant No. NNX08AR22G issued through the Planetary Science Division of the NASA Science Mission Directorate, the National Science Foundation grant No. AST-1238877, the University of Maryland, Eotvos Lorand University (ELTE), the Los Alamos National Laboratory, and the Gordon and Betty Moore Foundation.

GAMA is a joint European-Australasian project based around a spectroscopic campaign using the Anglo-Australian Telescope. The GAMA input catalogue is based on data taken from the Sloan Digital Sky Survey and the UKIRT Infrared Deep Sky Survey. Complementary imaging of the GAMA regions is being obtained by a number of independent survey programmes including GALEX MIS, VST KiDS, VISTA VIKING, WISE, Herschel-ATLAS, GMRT and ASKAP providing UV to radio coverage. GAMA is funded by the STFC (UK), the ARC (Australia), the AAO, and the participating institutions. The GAMA website is http://www.gama-survey.org/. 

Funding for SDSS-III has been provided by the Alfred P. Sloan Foundation, the Participating Institutions, the National Science Foundation, and the U.S. Department of Energy Office of Science. The SDSS-III web site is http://www.sdss3.org/. SDSS-III is managed by the Astrophysical Research Consortium for the Participating Institutions of the SDSS-III Collaboration including the University of Arizona, the Brazilian Participation Group, Brookhaven National Laboratory, University of Cambridge, Carnegie Mellon University, University of Florida, the French Participation Group, the German Participation Group, Harvard University, the Instituto de Astrofisica de Canarias, the Michigan State/Notre Dame/JINA Participation Group, Johns Hopkins University, Lawrence Berkeley National Laboratory, Max Planck Institute for Astrophysics, Max Planck Institute for Extraterrestrial Physics, New Mexico State University, New York University, Ohio State University, Pennsylvania State University, University of Portsmouth, Princeton University, the Spanish Participation Group, University of Tokyo, University of Utah, Vanderbilt University, University of Virginia, University of Washington, and Yale University.  

\software{Astropy \citep{astropy:2013, astropy:2018}, matplotlib \citep{Hunter:2007}, SciPy \citep{jones_scipy_2001}, the IPython package \citep{PER-GRA:2007}, NumPy \citep{van2011numpy}}

\appendix

\rrr{\section{Tests of the Background Subtraction}\label{s:backgroundsubtraction}}
\rrr{The fidelity of the background subtraction is of key importance 
for two aspects of our analysis. First, background subtraction schemes that
measure the sky background on scales comparable to the size of the target galaxies
are liable to attribute diffuse light in the target outskirts to the sky background.
This causes oversubtracted ``dark rings'' around galaxies that are large on the sky. 
Second, a uniform under- or oversubtraction of the background could affect our measurement
of \reff{}, as well as the estimate of our surface brightness limit.}

\rrr{
Dark rings were a significant problem for nearby massive galaxies 
in the first data release of the HSC-SSP, prompting
a change in the background estimation algorithm used for S18A (and the second data release).
The updated algorithm is described in detail in Section 4.1 of \citet{aihara2019}; here
we summarize the changes that most strongly affect performance for our sample.
}

\rrr{The most pertinent change to the sky subtraction algorithm of S18A (or equivalently,
PDR2) is that the sky is now estimated over the full focal plane, rather than over 
individual CCDs. In both cases, the sky is estimated by fitting a sixth-order
two-dimensional Chebyshev polynomial to ``superpixels'', which are themselves defined as
the clipped mean of non-detection pixels within an $N \times N$ pixel grid. However,
because the S18A pipeline fits a sky background over the entire focal plane of HSC, the
superpixels can be much larger ($N=1024$ pixels, $2'.8$ deg) than is possible with 
individual CCD sky estimation ($N=256$ pixels, $43''$). The maximum radial range at which
we measure the 1D surface brightness profiles in this work is 500 pixels (84''), 
significantly smaller than the S18A superpixels. Oversubtracted dark rings should
thus not occur around our target galaxies in S18A -- indeed, the galaxies are sufficiently
small on the sky that the dark ring oversubtraction was not likely to be present 
even in the PDR1 pipeline.}

\rrr{To quantitatively test this statement, as well as to estimate the overall residual
sky background in the vicinity of our targets, we follow the approach of Li et al. 2020,
in prep., who test the performance of the surface brightness sensitivity and S18A background
subtraction for a set of low-z ($z\sim0.02$) and intermediate-z ($z\sim0.40$) massive galaxies
in HSC. From tests on mock galaxies and comparisons to imaging from 
the Dragonfly Wide Field Survey \citep{danieli2020} and the Dark Energy Camera Legacy
Survey \citep[DECaLS, ][]{dey2019},
they conclude that the intermediate-redshift massive galaxies
are slightly uniformly undersubtracted, while the low-redshift massive galaxies are slightly 
uniformly oversubtracted. 
They moreover find that nearby `Sky Objects' (\textsf{SkyObj}), can be used to estimate and 
correct for the residual over- or undersubtraction of the sky background to reach
surface brightness sensitivities of $\mu_r \sim 29.5$ mag arcsec$^{-2}$. These Sky Objects are
identified by the HSC data reduction pipeline (see Section 6.6.8 of \citealt{aihara2019}) to
be locations in which no objects are detected. Sky Object photometry is
measured in apertures ranging from 20 arcsec
to 118 arcsec in diameter.}

\rrr{We estimate the residual background around our dwarf sample by taking the mean flux in
sky objects in the vicinity of our sample, as is done by Li et al. 2020, in prep for their
more massive galaxy sample. We expect that the sky around our dwarf galaxies will be
better estimated than the sky around the low-z massive galaxy for two reasons. First, our 
sample is characterized by nearly exponential profiles (\sersic{} indices close to $n=1$), meaning
that their surface brightness profiles drop more steeply with distance than do the
higher $n$ massive galaxies in Li et al. 2020, in prep. Second, the massive galaxies are much more
physically extended than our sample, and thus appear larger on-sky at fixed redshift.}

\begin{figure}[htb]
\centering     
\includegraphics[width=.5\linewidth]{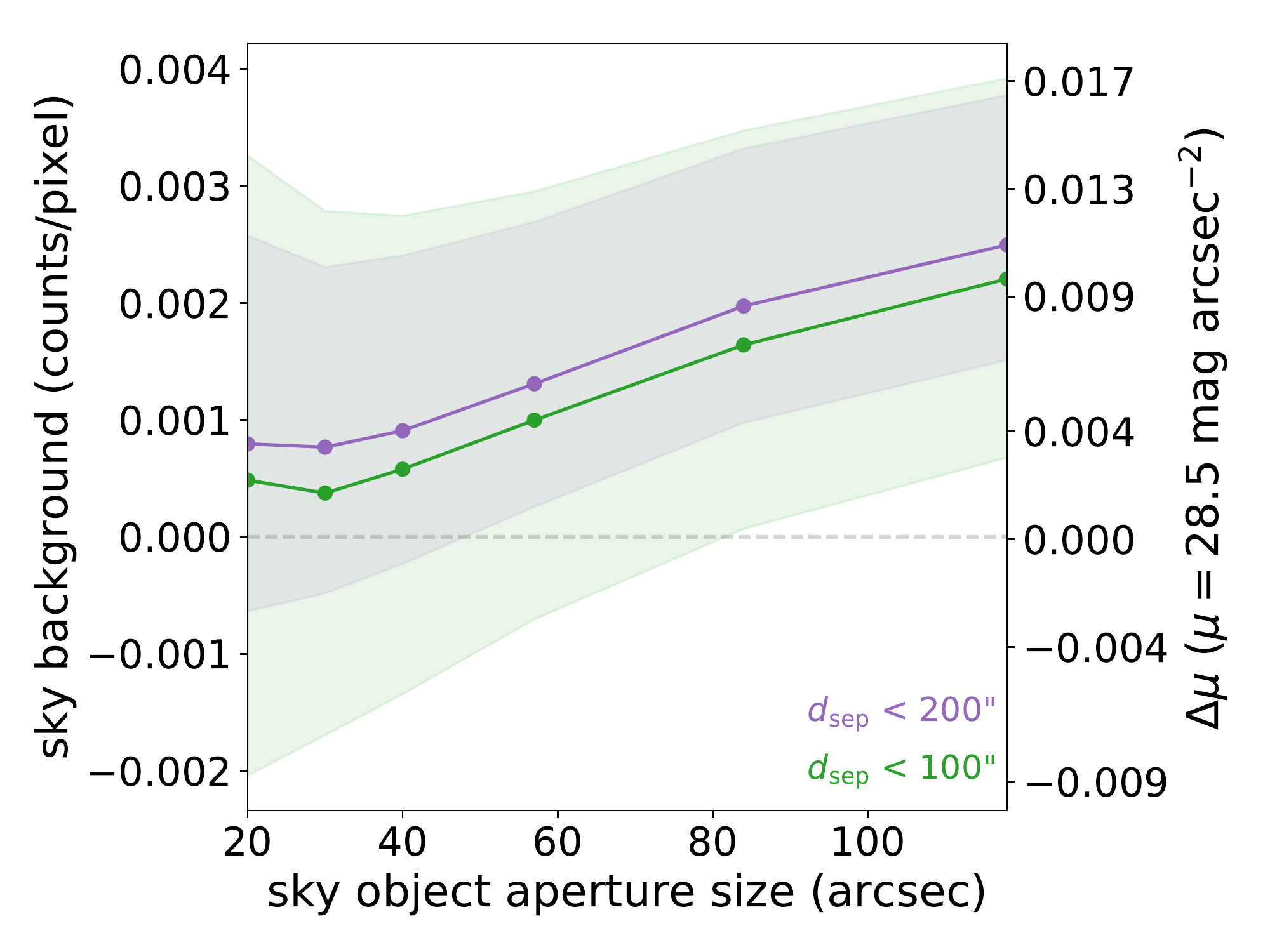}
\caption{ \rrr{The residual sky in the HSC-i band near our sample as 
measured from nearby Sky Objects in the vicinity of our target galaxies. The sky background
is measured as a function of Sky Object aperture size for Sky Objects within 100" (green, 43149
Sky Objects) and 200"
(purple, 172974 Sky Objects) of our target galaxy sample. 
At left, we give the sky background in counts/pixel. At right,
we show the change in surface brightness expected for an object with surface brightness
28.5 mag arcsec$^{-2}$. The solid curves show the median sky residual, while the shaded regions
show the 25\thh{} and 75\thh{} percentiles.
We find that the
residual background has a negligible impact ($\lesssim 0.02$ mag arcsec$^{-2}$) 
on the surface brightness measured at our nominal
surface brightness limit of 28.5 mag arcsec$^{-2}$. } }
\label{f:skybackground}
\end{figure}

\rrr{Indeed, we find that the background around our galaxies tends to be slightly undersubtracted
to a similar degree as the intermediate-redshift sample of Li et al. 2020, in prep. In
\autoref{f:skybackground} we show the estimate of the sky background as a function of Sky Object
aperture size for Sky Objects within 100 arcsec (green) and 200 arcsec (purple) of our target galaxies. 
For context, we reiterate that the maximum distance at which we measure profiles is 84 arcsec. 
We note that the two largest Sky Objects (of size 84 arcsec and 118 arcsec) 
are as large or larger than the extent over which we measure 1D profiles. Sky Objects with 
increasingly large apertures are more likely to include contributions from background objects;
The increase in the background level as a function of aperture size can be partially attributed to
this effect. 
Regardless, we find that in all cases the additive background residual 
is $\lesssim 0.004$ counts per pixel. 
This corresponds to a change in surface brightness of $\lesssim 0.02$ mag arcsec$^{-2}$ for
an object with surface brightness 28.5 mag arcsec$^{-2}$. The impact of the sky residual is small down to our nominal surface brightness limit
of 28.5 mag arcsec$^{-2}$. 
}

\section{Model Selection: Allowing for Shape-Size Covariance}\label{s:shapesize}
For a sample of galaxies at higher stellar mass and redshift, \cite{zhang2019} 
showed that the inferred 3D shape distribution changes significantly when 
allowing a non-zero covariance between galaxy size and ellipticity.

To test whether this model is necessary for our sample of dwarfs, we fit a multivariate
normal described by $\vec \alpha = \{ \mu_A, \mu_B, \mu_C, \sigma_{AA}, \sigma_{BB}, 
\sigma_{CC}, \sigma_{AC} \}$\footnote{here, $A$ is in kpc, whereas $B$ 
and $C$ are defined as normalized 
lengths relative to $A$}. As in \cite{zhang2019}, we only allow for a non-zero covariance
between $A$ and $C$. In order to understand whether the (${\rm R_{eff}}$, b/a) data demand
a non-zero shape-size covariance, we initiate walkers normally distributed about the 
best-fit values inferred for the fiducial models. For each parameter $X$, initial values
for the walkers are drawn from a normal distribution $\mathcal{N}(\tilde X, 0.25\tilde X)$ where
$\tilde X$ is the best-fit value from the fiducial model.

\begin{figure}[htb]
\centering     
\includegraphics[width=\linewidth]{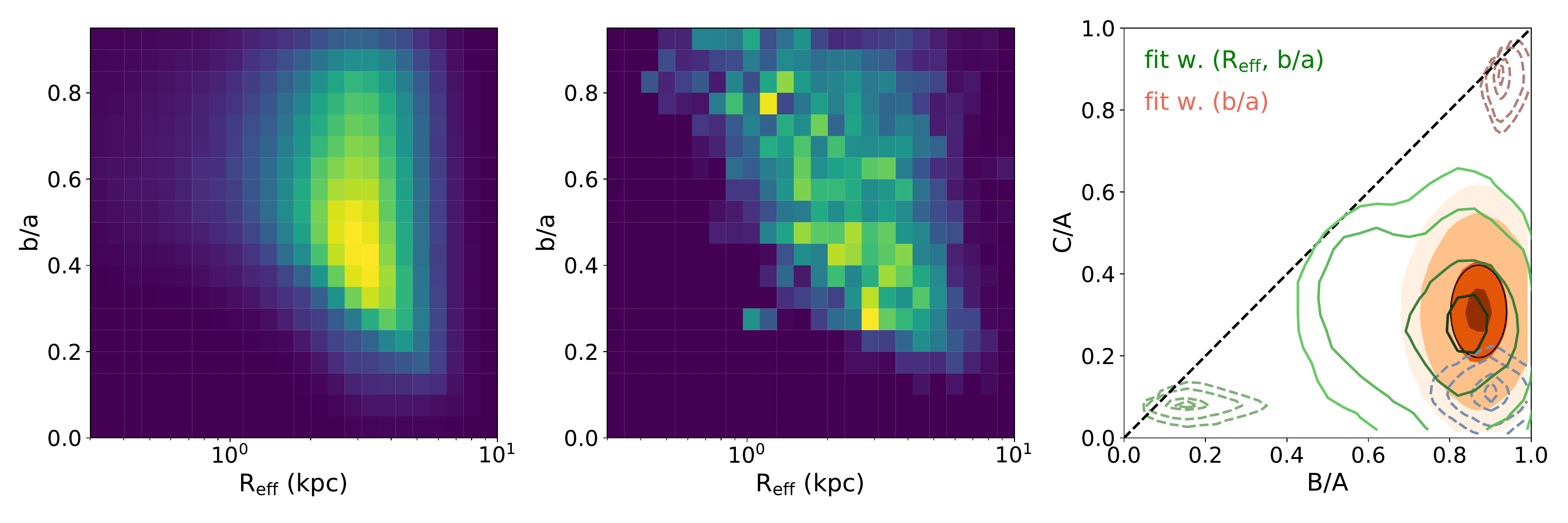}
\caption{\textit{Left:} the distribution of effective radius and projected
axis ratio for a sample drawn from the posterior distribution for a 
model where we allow for a non-zero covariance between $C/A$ and
\reff{}. \textit{Middle:} the distribution over \reff{} and projected
axis ratio ($b/a$) for the observed galaxies. \textit{Right:} the 
inferred distribution over intrinsic axis ratios for the fiducial
model (orange filled contours) and the model at left (green unfilled
contours).
    }
\label{f:logainference}
\end{figure}

The results of this test are shown in \autoref{f:logainference}. Though the data are
well-described by a slightly negative shape-size covariance $\sigma_{AC} \sim -0.16$, we find
that this model yields a small shift towards lower $B/A$, and
a negligible shift in $C/A$. Therefore, because the model of
\citep{zhang2019} adds three additional degrees of freedom to the model ($\mu_A$, $\sigma_A$,
and $\sigma_{AC}$), we choose to fit the marginalized b/a distribution, rather than 
$({\rm R_{eff}},$b/a).

\section{Parameter Recovery and Sample Size}\label{s:samplesize}
\begin{figure*}[htb]
\centering     
\includegraphics[width=\linewidth]{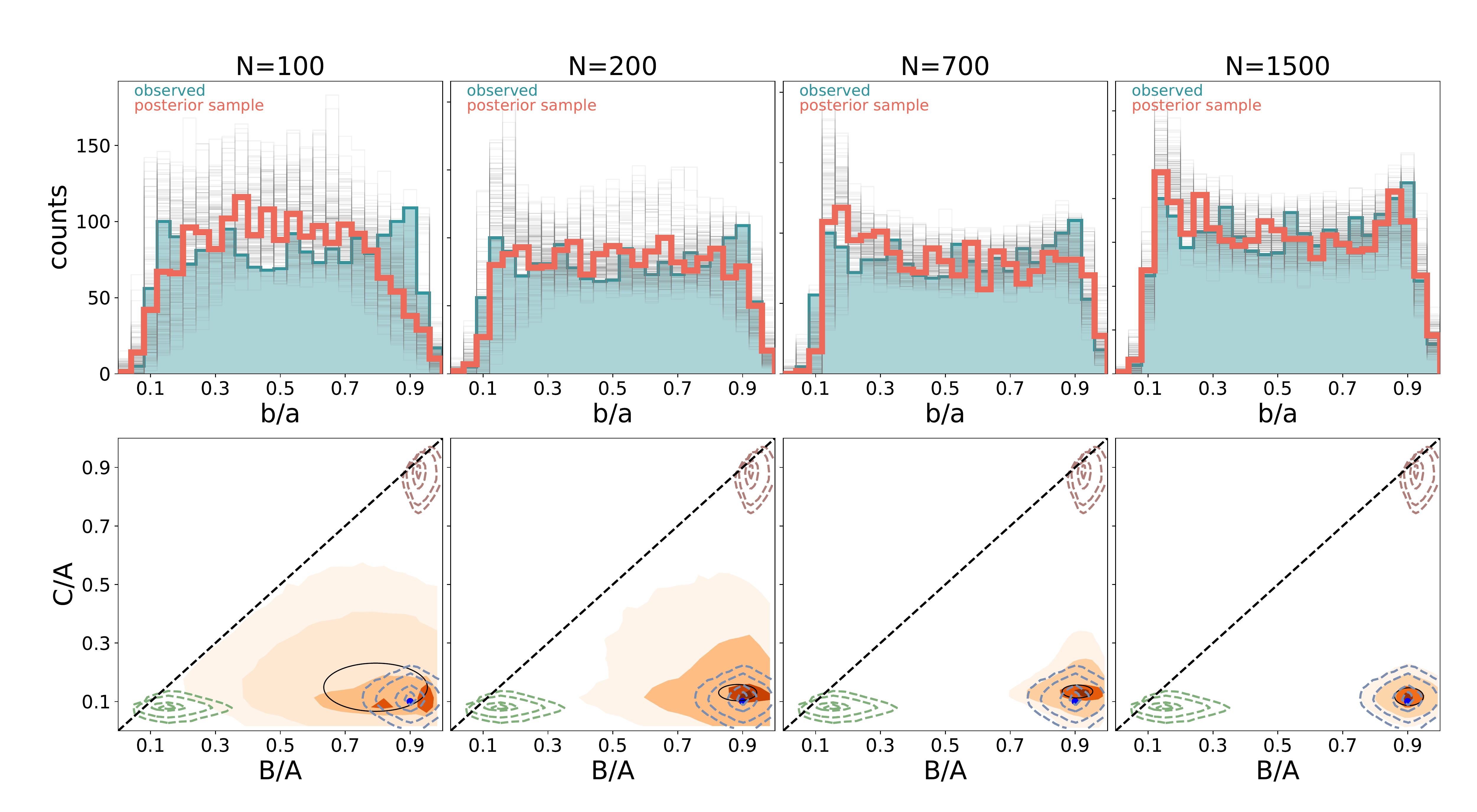}
\caption{ 
For a population of mock disk galaxies described by 
$\vec\alpha=\{\mu_B,\mu_C,\sigma_B,\sigma_C\}=\{0.9,0.1,0.,0.\}$, 
we investigate the precision at which the model parameters $\vec\alpha$
may be recovered as a function of sample size $N$. The top row
shows, in each panel, the true distribution of the projected axis ratio
$b/a$ in teal and the posterior sample in orange. Individual draws from
the posterior are shown in grey. The bottom row shows the distribution
of intrinsic axis ratios ($B/A$ and $C/A$) from the posterior sample in
orange. The maximum a posteriori estimate is shown by the black ellipse.
The pure disk, prolate, and spheroidal populations are shown by 
dashed contours (see \autoref{f:boundaries}). A successful recovery is 
one in which the orange contours of the posterior samples coincides with
the dashed blue contours of the pure disk population.
    }
\label{f:sersicNrecovery}
\end{figure*}

As sample size decreases, the precision of our inferred model parameters
$\vec\alpha=\{\mu_B,\mu_C,\sigma_B,\sigma_C\}$ decreases in kind. 
In \autoref{f:sersicNrecovery}, we show the distribution of
intrinsic axis ratios inferred for a sample of N=\{100,200,1500,2500\} 
galaxies. We find that at N$\sim200$ galaxies, though 
$\mu_B$ and $\mu_C$ are well-recovered, $\sigma_B$ and $\sigma_C$ are
increasingly overestimated\footnote{We note that $\sigma_B$ and $\sigma_C$
are always overestimated when $\sigma_B=\sigma_C=0$ due to the 
uncertainty in measuring 1D surface brightness profiles}. We thus
choose to only consider subsets of galaxies where N$>700$. 

{}

\bibliography{dwarfshapes.bib}

\end{document}